\documentclass[longauth]{aaEC}

\usepackage{graphicx}
\usepackage{natbib}
\usepackage{scalerel}
\usepackage{xspace}
\usepackage[table]{xcolor}
\usepackage{asymptote}

\usepackage{txfonts}

\usepackage[pdfencoding=auto,psdextra]{hyperref}
\hypersetup{
    colorlinks=true,
    linkcolor=blue,
    filecolor=magenta,      
    urlcolor=blue,
    citecolor=blue
}
\makeatletter
\renewcommand*\aa@pageof{, page \thepage{} of \pageref*{LastPage}}
\makeatother

\usepackage[utf8]{inputenc}

\usepackage[switch, modulo]{lineno}

\usepackage{euclid}

\begin{document}

%
%

\title{\Euclid preparation. 3-dimensional galaxy clustering in configuration space Part \textrm{I}: 2-point correlation function estimation}



\newcommand{\nc}{\newcommand}
\nc{\br}{\Vec{r}}
\nc{\bx}{\Vec{x}}
\nc{\rpe}{r_\perp}
\nc{\rpa}{{r_\parallel}} 
\nc{\xir}{\xi(r)}
\nc{\xirv}{\xi(\br)}
\nc{\xirppi}{\xi(\rpe,\rpa)}
\nc{\xismu}{\xi(r,\mu)}
\nc{\ximono}{\xi_0(r)}
\nc{\xiquad}{\xi_2(r)}
\nc{\xihexa}{\xi_4(r)}
\nc{\wprp}{w_\perp(\rpe)}

\newcommand{\orcid}[1]{} 
\author{Euclid Collaboration: S.~de~la~Torre\thanks{\email{sylvain.delatorre@lam.fr}}\inst{\ref{aff1}}
\and F.~Marulli\orcid{0000-0002-8850-0303}\inst{\ref{aff2},\ref{aff3},\ref{aff4}}
\and E.~Keih\"anen\orcid{0000-0003-1804-7715}\inst{\ref{aff5}}
\and A.~Viitanen\orcid{0000-0001-9383-786X}\inst{\ref{aff5},\ref{aff6}}
\and M.~Viel\orcid{0000-0002-2642-5707}\inst{\ref{aff7},\ref{aff8},\ref{aff9},\ref{aff10},\ref{aff11}}
\and A.~Veropalumbo\orcid{0000-0003-2387-1194}\inst{\ref{aff12},\ref{aff13},\ref{aff14}}
\and E.~Branchini\orcid{0000-0002-0808-6908}\inst{\ref{aff15},\ref{aff13},\ref{aff12}}
\and D.~Tavagnacco\orcid{0000-0001-7475-9894}\inst{\ref{aff8}}
\and F.~Rizzo\orcid{0000-0002-9407-585X}\inst{\ref{aff8}}
\and J.~Valiviita\orcid{0000-0001-6225-3693}\inst{\ref{aff16},\ref{aff17}}
\and V.~Lindholm\orcid{0000-0003-2317-5471}\inst{\ref{aff16},\ref{aff17}}
\and V.~Allevato\orcid{0000-0001-7232-5152}\inst{\ref{aff18}}
\and G.~Parimbelli\orcid{0000-0002-2539-2472}\inst{\ref{aff19},\ref{aff14},\ref{aff9}}
\and E.~Sarpa\orcid{0000-0002-1256-655X}\inst{\ref{aff9},\ref{aff11},\ref{aff10}}
\and Z.~Ghaffari\orcid{0000-0002-6467-8078}\inst{\ref{aff8},\ref{aff7}}
\and A.~Amara\inst{\ref{aff20}}
\and S.~Andreon\orcid{0000-0002-2041-8784}\inst{\ref{aff12}}
\and N.~Auricchio\orcid{0000-0003-4444-8651}\inst{\ref{aff3}}
\and C.~Baccigalupi\orcid{0000-0002-8211-1630}\inst{\ref{aff7},\ref{aff8},\ref{aff10},\ref{aff9}}
\and M.~Baldi\orcid{0000-0003-4145-1943}\inst{\ref{aff21},\ref{aff3},\ref{aff4}}
\and S.~Bardelli\orcid{0000-0002-8900-0298}\inst{\ref{aff3}}
\and A.~Basset\inst{\ref{aff22}}
\and D.~Bonino\orcid{0000-0002-3336-9977}\inst{\ref{aff23}}
\and M.~Brescia\orcid{0000-0001-9506-5680}\inst{\ref{aff24},\ref{aff18},\ref{aff25}}
\and J.~Brinchmann\orcid{0000-0003-4359-8797}\inst{\ref{aff26},\ref{aff27}}
\and A.~Caillat\inst{\ref{aff1}}
\and S.~Camera\orcid{0000-0003-3399-3574}\inst{\ref{aff28},\ref{aff29},\ref{aff23}}
\and V.~Capobianco\orcid{0000-0002-3309-7692}\inst{\ref{aff23}}
\and C.~Carbone\orcid{0000-0003-0125-3563}\inst{\ref{aff30}}
\and J.~Carretero\orcid{0000-0002-3130-0204}\inst{\ref{aff31},\ref{aff32}}
\and S.~Casas\orcid{0000-0002-4751-5138}\inst{\ref{aff33},\ref{aff34}}
\and F.~J.~Castander\orcid{0000-0001-7316-4573}\inst{\ref{aff19},\ref{aff35}}
\and M.~Castellano\orcid{0000-0001-9875-8263}\inst{\ref{aff6}}
\and G.~Castignani\orcid{0000-0001-6831-0687}\inst{\ref{aff3}}
\and S.~Cavuoti\orcid{0000-0002-3787-4196}\inst{\ref{aff18},\ref{aff25}}
\and A.~Cimatti\inst{\ref{aff36}}
\and C.~Colodro-Conde\inst{\ref{aff37}}
\and G.~Congedo\orcid{0000-0003-2508-0046}\inst{\ref{aff38}}
\and C.~J.~Conselice\orcid{0000-0003-1949-7638}\inst{\ref{aff39}}
\and L.~Conversi\orcid{0000-0002-6710-8476}\inst{\ref{aff40},\ref{aff41}}
\and Y.~Copin\orcid{0000-0002-5317-7518}\inst{\ref{aff42}}
\and F.~Courbin\orcid{0000-0003-0758-6510}\inst{\ref{aff43},\ref{aff44}}
\and H.~M.~Courtois\orcid{0000-0003-0509-1776}\inst{\ref{aff45}}
\and M.~Crocce\orcid{0000-0002-9745-6228}\inst{\ref{aff19},\ref{aff35}}
\and A.~Da~Silva\orcid{0000-0002-6385-1609}\inst{\ref{aff46},\ref{aff47}}
\and H.~Degaudenzi\orcid{0000-0002-5887-6799}\inst{\ref{aff48}}
\and G.~De~Lucia\orcid{0000-0002-6220-9104}\inst{\ref{aff8}}
\and A.~M.~Di~Giorgio\orcid{0000-0002-4767-2360}\inst{\ref{aff49}}
\and J.~Dinis\orcid{0000-0001-5075-1601}\inst{\ref{aff46},\ref{aff47}}
\and F.~Dubath\orcid{0000-0002-6533-2810}\inst{\ref{aff48}}
\and C.~A.~J.~Duncan\orcid{0009-0003-3573-0791}\inst{\ref{aff39}}
\and X.~Dupac\inst{\ref{aff41}}
\and S.~Dusini\orcid{0000-0002-1128-0664}\inst{\ref{aff50}}
\and M.~Farina\orcid{0000-0002-3089-7846}\inst{\ref{aff49}}
\and S.~Farrens\orcid{0000-0002-9594-9387}\inst{\ref{aff51}}
\and F.~Faustini\orcid{0000-0001-6274-5145}\inst{\ref{aff52},\ref{aff6}}
\and S.~Ferriol\inst{\ref{aff42}}
\and N.~Fourmanoit\orcid{0009-0005-6816-6925}\inst{\ref{aff53}}
\and M.~Frailis\orcid{0000-0002-7400-2135}\inst{\ref{aff8}}
\and E.~Franceschi\orcid{0000-0002-0585-6591}\inst{\ref{aff3}}
\and P.~Franzetti\inst{\ref{aff30}}
\and M.~Fumana\orcid{0000-0001-6787-5950}\inst{\ref{aff30}}
\and S.~Galeotta\orcid{0000-0002-3748-5115}\inst{\ref{aff8}}
\and K.~George\orcid{0000-0002-1734-8455}\inst{\ref{aff54}}
\and W.~Gillard\orcid{0000-0003-4744-9748}\inst{\ref{aff53}}
\and B.~Gillis\orcid{0000-0002-4478-1270}\inst{\ref{aff38}}
\and C.~Giocoli\orcid{0000-0002-9590-7961}\inst{\ref{aff3},\ref{aff4}}
\and P.~G\'omez-Alvarez\orcid{0000-0002-8594-5358}\inst{\ref{aff55},\ref{aff41}}
\and B.~R.~Granett\orcid{0000-0003-2694-9284}\inst{\ref{aff12}}
\and A.~Grazian\orcid{0000-0002-5688-0663}\inst{\ref{aff56}}
\and F.~Grupp\inst{\ref{aff57},\ref{aff54}}
\and L.~Guzzo\orcid{0000-0001-8264-5192}\inst{\ref{aff58},\ref{aff12}}
\and S.~V.~H.~Haugan\orcid{0000-0001-9648-7260}\inst{\ref{aff59}}
\and W.~Holmes\inst{\ref{aff60}}
\and F.~Hormuth\inst{\ref{aff61}}
\and A.~Hornstrup\orcid{0000-0002-3363-0936}\inst{\ref{aff62},\ref{aff63}}
\and S.~Ili\'c\orcid{0000-0003-4285-9086}\inst{\ref{aff64},\ref{aff65}}
\and K.~Jahnke\orcid{0000-0003-3804-2137}\inst{\ref{aff66}}
\and M.~Jhabvala\inst{\ref{aff67}}
\and B.~Joachimi\orcid{0000-0001-7494-1303}\inst{\ref{aff68}}
\and S.~Kermiche\orcid{0000-0002-0302-5735}\inst{\ref{aff53}}
\and A.~Kiessling\orcid{0000-0002-2590-1273}\inst{\ref{aff60}}
\and M.~Kilbinger\orcid{0000-0001-9513-7138}\inst{\ref{aff51}}
\and B.~Kubik\orcid{0009-0006-5823-4880}\inst{\ref{aff42}}
\and M.~Kunz\orcid{0000-0002-3052-7394}\inst{\ref{aff69}}
\and H.~Kurki-Suonio\orcid{0000-0002-4618-3063}\inst{\ref{aff16},\ref{aff17}}
\and S.~Ligori\orcid{0000-0003-4172-4606}\inst{\ref{aff23}}
\and P.~B.~Lilje\orcid{0000-0003-4324-7794}\inst{\ref{aff59}}
\and I.~Lloro\orcid{0000-0001-5966-1434}\inst{\ref{aff70}}
\and G.~Mainetti\orcid{0000-0003-2384-2377}\inst{\ref{aff71}}
\and D.~Maino\inst{\ref{aff58},\ref{aff30},\ref{aff72}}
\and E.~Maiorano\orcid{0000-0003-2593-4355}\inst{\ref{aff3}}
\and O.~Mansutti\orcid{0000-0001-5758-4658}\inst{\ref{aff8}}
\and O.~Marggraf\orcid{0000-0001-7242-3852}\inst{\ref{aff73}}
\and K.~Markovic\orcid{0000-0001-6764-073X}\inst{\ref{aff60}}
\and M.~Martinelli\orcid{0000-0002-6943-7732}\inst{\ref{aff6},\ref{aff74}}
\and N.~Martinet\orcid{0000-0003-2786-7790}\inst{\ref{aff1}}
\and R.~Massey\orcid{0000-0002-6085-3780}\inst{\ref{aff75}}
\and S.~Maurogordato\inst{\ref{aff76}}
\and E.~Medinaceli\orcid{0000-0002-4040-7783}\inst{\ref{aff3}}
\and S.~Mei\orcid{0000-0002-2849-559X}\inst{\ref{aff77}}
\and M.~MELMhior\inst{\ref{aff78}}
\and Y.~Mellier\inst{\ref{aff79},\ref{aff80}}
\and M.~Meneghetti\orcid{0000-0003-1225-7084}\inst{\ref{aff3},\ref{aff4}}
\and E.~Merlin\orcid{0000-0001-6870-8900}\inst{\ref{aff6}}
\and G.~Meylan\inst{\ref{aff81}}
\and M.~Moresco\orcid{0000-0002-7616-7136}\inst{\ref{aff2},\ref{aff3}}
\and B.~Morin\inst{\ref{aff51}}
\and L.~Moscardini\orcid{0000-0002-3473-6716}\inst{\ref{aff2},\ref{aff3},\ref{aff4}}
\and E.~Munari\orcid{0000-0002-1751-5946}\inst{\ref{aff8},\ref{aff7}}
\and C.~Neissner\orcid{0000-0001-8524-4968}\inst{\ref{aff82},\ref{aff32}}
\and S.-M.~Niemi\inst{\ref{aff83}}
\and C.~Padilla\orcid{0000-0001-7951-0166}\inst{\ref{aff82}}
\and S.~Paltani\orcid{0000-0002-8108-9179}\inst{\ref{aff48}}
\and F.~Pasian\orcid{0000-0002-4869-3227}\inst{\ref{aff8}}
\and K.~Pedersen\inst{\ref{aff84}}
\and W.~J.~Percival\orcid{0000-0002-0644-5727}\inst{\ref{aff85},\ref{aff86},\ref{aff87}}
\and V.~Pettorino\inst{\ref{aff83}}
\and S.~Pires\orcid{0000-0002-0249-2104}\inst{\ref{aff51}}
\and G.~Polenta\orcid{0000-0003-4067-9196}\inst{\ref{aff52}}
\and M.~Poncet\inst{\ref{aff22}}
\and L.~Pozzetti\orcid{0000-0001-7085-0412}\inst{\ref{aff3}}
\and F.~Raison\orcid{0000-0002-7819-6918}\inst{\ref{aff57}}
\and A.~Renzi\orcid{0000-0001-9856-1970}\inst{\ref{aff88},\ref{aff50}}
\and J.~Rhodes\orcid{0000-0002-4485-8549}\inst{\ref{aff60}}
\and G.~Riccio\inst{\ref{aff18}}
\and E.~Romelli\orcid{0000-0003-3069-9222}\inst{\ref{aff8}}
\and M.~Roncarelli\orcid{0000-0001-9587-7822}\inst{\ref{aff3}}
\and E.~Rossetti\orcid{0000-0003-0238-4047}\inst{\ref{aff21}}
\and R.~Saglia\orcid{0000-0003-0378-7032}\inst{\ref{aff54},\ref{aff57}}
\and Z.~Sakr\orcid{0000-0002-4823-3757}\inst{\ref{aff89},\ref{aff65},\ref{aff90}}
\and A.~G.~S\'anchez\orcid{0000-0003-1198-831X}\inst{\ref{aff57}}
\and D.~Sapone\orcid{0000-0001-7089-4503}\inst{\ref{aff91}}
\and B.~Sartoris\orcid{0000-0003-1337-5269}\inst{\ref{aff54},\ref{aff8}}
\and P.~Schneider\orcid{0000-0001-8561-2679}\inst{\ref{aff73}}
\and T.~Schrabback\orcid{0000-0002-6987-7834}\inst{\ref{aff92}}
\and M.~Scodeggio\inst{\ref{aff30}}
\and A.~Secroun\orcid{0000-0003-0505-3710}\inst{\ref{aff53}}
\and E.~Sefusatti\orcid{0000-0003-0473-1567}\inst{\ref{aff8},\ref{aff7},\ref{aff10}}
\and G.~Seidel\orcid{0000-0003-2907-353X}\inst{\ref{aff66}}
\and M.~Seiffert\orcid{0000-0002-7536-9393}\inst{\ref{aff60}}
\and S.~Serrano\orcid{0000-0002-0211-2861}\inst{\ref{aff35},\ref{aff93},\ref{aff19}}
\and C.~Sirignano\orcid{0000-0002-0995-7146}\inst{\ref{aff88},\ref{aff50}}
\and G.~Sirri\orcid{0000-0003-2626-2853}\inst{\ref{aff4}}
\and L.~Stanco\orcid{0000-0002-9706-5104}\inst{\ref{aff50}}
\and J.~Steinwagner\orcid{0000-0001-7443-1047}\inst{\ref{aff57}}
\and C.~Surace\orcid{0000-0003-2592-0113}\inst{\ref{aff1}}
\and P.~Tallada-Cresp\'{i}\orcid{0000-0002-1336-8328}\inst{\ref{aff31},\ref{aff32}}
\and A.~N.~Taylor\inst{\ref{aff38}}
\and I.~Tereno\inst{\ref{aff46},\ref{aff94}}
\and R.~Toledo-Moreo\orcid{0000-0002-2997-4859}\inst{\ref{aff95}}
\and F.~Torradeflot\orcid{0000-0003-1160-1517}\inst{\ref{aff32},\ref{aff31}}
\and A.~Tsyganov\inst{\ref{aff96}}
\and I.~Tutusaus\orcid{0000-0002-3199-0399}\inst{\ref{aff65}}
\and L.~Valenziano\orcid{0000-0002-1170-0104}\inst{\ref{aff3},\ref{aff97}}
\and T.~Vassallo\orcid{0000-0001-6512-6358}\inst{\ref{aff54},\ref{aff8}}
\and Y.~Wang\orcid{0000-0002-4749-2984}\inst{\ref{aff98}}
\and J.~Weller\orcid{0000-0002-8282-2010}\inst{\ref{aff54},\ref{aff57}}
\and A.~Zacchei\orcid{0000-0003-0396-1192}\inst{\ref{aff8},\ref{aff7}}
\and G.~Zamorani\orcid{0000-0002-2318-301X}\inst{\ref{aff3}}
\and E.~Zucca\orcid{0000-0002-5845-8132}\inst{\ref{aff3}}
\and A.~Biviano\orcid{0000-0002-0857-0732}\inst{\ref{aff8},\ref{aff7}}
\and M.~Bolzonella\orcid{0000-0003-3278-4607}\inst{\ref{aff3}}
\and E.~Bozzo\orcid{0000-0002-8201-1525}\inst{\ref{aff48}}
\and C.~Burigana\orcid{0000-0002-3005-5796}\inst{\ref{aff99},\ref{aff97}}
\and M.~Calabrese\orcid{0000-0002-2637-2422}\inst{\ref{aff100},\ref{aff30}}
\and D.~Di~Ferdinando\inst{\ref{aff4}}
\and J.~A.~Escartin~Vigo\inst{\ref{aff57}}
\and R.~Farinelli\inst{\ref{aff3}}
\and F.~Finelli\orcid{0000-0002-6694-3269}\inst{\ref{aff3},\ref{aff97}}
\and L.~Gabarra\orcid{0000-0002-8486-8856}\inst{\ref{aff101}}
\and J.~Gracia-Carpio\inst{\ref{aff57}}
\and S.~Matthew\orcid{0000-0001-8448-1697}\inst{\ref{aff38}}
\and N.~Mauri\orcid{0000-0001-8196-1548}\inst{\ref{aff36},\ref{aff4}}
\and A.~Mora\orcid{0000-0002-1922-8529}\inst{\ref{aff102}}
\and A.~Pezzotta\orcid{0000-0003-0726-2268}\inst{\ref{aff57}}
\and M.~P\"ontinen\orcid{0000-0001-5442-2530}\inst{\ref{aff16}}
\and V.~Scottez\inst{\ref{aff79},\ref{aff103}}
\and P.~Simon\inst{\ref{aff73}}
\and A.~Spurio~Mancini\orcid{0000-0001-5698-0990}\inst{\ref{aff104},\ref{aff105}}
\and M.~Tenti\orcid{0000-0002-4254-5901}\inst{\ref{aff4}}
\and M.~Wiesmann\orcid{0009-0000-8199-5860}\inst{\ref{aff59}}
\and Y.~Akrami\orcid{0000-0002-2407-7956}\inst{\ref{aff106},\ref{aff107}}
\and I.~T.~Andika\orcid{0000-0001-6102-9526}\inst{\ref{aff108},\ref{aff109}}
\and S.~Anselmi\orcid{0000-0002-3579-9583}\inst{\ref{aff50},\ref{aff88},\ref{aff110}}
\and M.~Archidiacono\orcid{0000-0003-4952-9012}\inst{\ref{aff58},\ref{aff72}}
\and F.~Atrio-Barandela\orcid{0000-0002-2130-2513}\inst{\ref{aff111}}
\and A.~Balaguera-Antolinez\orcid{0000-0001-5028-3035}\inst{\ref{aff37},\ref{aff112}}
\and D.~Bertacca\orcid{0000-0002-2490-7139}\inst{\ref{aff88},\ref{aff56},\ref{aff50}}
\and M.~Bethermin\orcid{0000-0002-3915-2015}\inst{\ref{aff113},\ref{aff1}}
\and A.~Blanchard\orcid{0000-0001-8555-9003}\inst{\ref{aff65}}
\and L.~Blot\orcid{0000-0002-9622-7167}\inst{\ref{aff114},\ref{aff110}}
\and H.~B\"ohringer\orcid{0000-0001-8241-4204}\inst{\ref{aff57},\ref{aff115},\ref{aff116}}
\and S.~Borgani\orcid{0000-0001-6151-6439}\inst{\ref{aff117},\ref{aff7},\ref{aff8},\ref{aff10},\ref{aff11}}
\and M.~L.~Brown\orcid{0000-0002-0370-8077}\inst{\ref{aff39}}
\and S.~Bruton\orcid{0000-0002-6503-5218}\inst{\ref{aff118}}
\and R.~Cabanac\orcid{0000-0001-6679-2600}\inst{\ref{aff65}}
\and A.~Calabro\orcid{0000-0003-2536-1614}\inst{\ref{aff6}}
\and B.~Camacho~Quevedo\orcid{0000-0002-8789-4232}\inst{\ref{aff35},\ref{aff19}}
\and G.~Ca\~nas-Herrera\orcid{0000-0003-2796-2149}\inst{\ref{aff83},\ref{aff119}}
\and A.~Cappi\inst{\ref{aff3},\ref{aff76}}
\and F.~Caro\inst{\ref{aff6}}
\and C.~S.~Carvalho\inst{\ref{aff94}}
\and T.~Castro\orcid{0000-0002-6292-3228}\inst{\ref{aff8},\ref{aff10},\ref{aff7},\ref{aff11}}
\and K.~C.~Chambers\orcid{0000-0001-6965-7789}\inst{\ref{aff120}}
\and F.~Cogato\orcid{0000-0003-4632-6113}\inst{\ref{aff2},\ref{aff3}}
\and S.~Contarini\orcid{0000-0002-9843-723X}\inst{\ref{aff57}}
\and A.~R.~Cooray\orcid{0000-0002-3892-0190}\inst{\ref{aff121}}
\and O.~Cucciati\orcid{0000-0002-9336-7551}\inst{\ref{aff3}}
\and S.~Davini\orcid{0000-0003-3269-1718}\inst{\ref{aff13}}
\and F.~De~Paolis\orcid{0000-0001-6460-7563}\inst{\ref{aff122},\ref{aff123},\ref{aff124}}
\and G.~Desprez\orcid{0000-0001-8325-1742}\inst{\ref{aff125}}
\and A.~D\'iaz-S\'anchez\orcid{0000-0003-0748-4768}\inst{\ref{aff126}}
\and S.~Di~Domizio\orcid{0000-0003-2863-5895}\inst{\ref{aff15},\ref{aff13}}
\and H.~Dole\orcid{0000-0002-9767-3839}\inst{\ref{aff127}}
\and S.~Escoffier\orcid{0000-0002-2847-7498}\inst{\ref{aff53}}
\and A.~G.~Ferrari\orcid{0009-0005-5266-4110}\inst{\ref{aff4}}
\and P.~G.~Ferreira\orcid{0000-0002-3021-2851}\inst{\ref{aff101}}
\and A.~Finoguenov\orcid{0000-0002-4606-5403}\inst{\ref{aff16}}
\and A.~Fontana\orcid{0000-0003-3820-2823}\inst{\ref{aff6}}
\and K.~Ganga\orcid{0000-0001-8159-8208}\inst{\ref{aff77}}
\and J.~Garc\'ia-Bellido\orcid{0000-0002-9370-8360}\inst{\ref{aff106}}
\and T.~Gasparetto\orcid{0000-0002-7913-4866}\inst{\ref{aff8}}
\and V.~Gautard\inst{\ref{aff128}}
\and E.~Gaztanaga\orcid{0000-0001-9632-0815}\inst{\ref{aff19},\ref{aff35},\ref{aff34}}
\and F.~Giacomini\orcid{0000-0002-3129-2814}\inst{\ref{aff4}}
\and F.~Gianotti\orcid{0000-0003-4666-119X}\inst{\ref{aff3}}
\and G.~Gozaliasl\orcid{0000-0002-0236-919X}\inst{\ref{aff129}}
\and A.~Gregorio\orcid{0000-0003-4028-8785}\inst{\ref{aff117},\ref{aff8},\ref{aff10}}
\and M.~Guidi\orcid{0000-0001-9408-1101}\inst{\ref{aff21},\ref{aff3}}
\and C.~M.~Gutierrez\orcid{0000-0001-7854-783X}\inst{\ref{aff130}}
\and A.~Hall\orcid{0000-0002-3139-8651}\inst{\ref{aff38}}
\and S.~Hemmati\orcid{0000-0003-2226-5395}\inst{\ref{aff131}}
\and H.~Hildebrandt\orcid{0000-0002-9814-3338}\inst{\ref{aff132}}
\and J.~Hjorth\orcid{0000-0002-4571-2306}\inst{\ref{aff84}}
\and A.~Jimenez~Mu\~noz\orcid{0009-0004-5252-185X}\inst{\ref{aff133}}
\and S.~Joudaki\orcid{0000-0001-8820-673X}\inst{\ref{aff34}}
\and J.~J.~E.~Kajava\orcid{0000-0002-3010-8333}\inst{\ref{aff134},\ref{aff135}}
\and Y.~Kang\orcid{0009-0000-8588-7250}\inst{\ref{aff48}}
\and V.~Kansal\orcid{0000-0002-4008-6078}\inst{\ref{aff136},\ref{aff137}}
\and D.~Karagiannis\orcid{0000-0002-4927-0816}\inst{\ref{aff138},\ref{aff139}}
\and C.~C.~Kirkpatrick\inst{\ref{aff5}}
\and S.~Kruk\orcid{0000-0001-8010-8879}\inst{\ref{aff41}}
\and M.~Lattanzi\orcid{0000-0003-1059-2532}\inst{\ref{aff140}}
\and A.~M.~C.~Le~Brun\orcid{0000-0002-0936-4594}\inst{\ref{aff110}}
\and S.~Lee\orcid{0000-0002-8289-740X}\inst{\ref{aff60}}
\and J.~Le~Graet\orcid{0000-0001-6523-7971}\inst{\ref{aff53}}
\and L.~Legrand\orcid{0000-0003-0610-5252}\inst{\ref{aff141},\ref{aff142}}
\and M.~Lembo\orcid{0000-0002-5271-5070}\inst{\ref{aff138},\ref{aff140}}
\and J.~Lesgourgues\orcid{0000-0001-7627-353X}\inst{\ref{aff33}}
\and T.~I.~Liaudat\orcid{0000-0002-9104-314X}\inst{\ref{aff143}}
\and A.~Loureiro\orcid{0000-0002-4371-0876}\inst{\ref{aff144},\ref{aff145}}
\and J.~Macias-Perez\orcid{0000-0002-5385-2763}\inst{\ref{aff133}}
\and M.~Magliocchetti\orcid{0000-0001-9158-4838}\inst{\ref{aff49}}
\and F.~Mannucci\orcid{0000-0002-4803-2381}\inst{\ref{aff146}}
\and R.~Maoli\orcid{0000-0002-6065-3025}\inst{\ref{aff147},\ref{aff6}}
\and J.~Mart\'{i}n-Fleitas\orcid{0000-0002-8594-569X}\inst{\ref{aff102}}
\and C.~J.~A.~P.~Martins\orcid{0000-0002-4886-9261}\inst{\ref{aff148},\ref{aff26}}
\and L.~Maurin\orcid{0000-0002-8406-0857}\inst{\ref{aff127}}
\and R.~B.~Metcalf\orcid{0000-0003-3167-2574}\inst{\ref{aff2},\ref{aff3}}
\and M.~Miluzio\inst{\ref{aff41},\ref{aff149}}
\and P.~Monaco\orcid{0000-0003-2083-7564}\inst{\ref{aff117},\ref{aff8},\ref{aff10},\ref{aff7}}
\and C.~Moretti\orcid{0000-0003-3314-8936}\inst{\ref{aff9},\ref{aff11},\ref{aff8},\ref{aff7},\ref{aff10}}
\and G.~Morgante\inst{\ref{aff3}}
\and C.~Murray\inst{\ref{aff77}}
\and S.~Nadathur\orcid{0000-0001-9070-3102}\inst{\ref{aff34}}
\and K.~Naidoo\orcid{0000-0002-9182-1802}\inst{\ref{aff34}}
\and A.~Navarro-Alsina\orcid{0000-0002-3173-2592}\inst{\ref{aff73}}
\and S.~Nesseris\orcid{0000-0002-0567-0324}\inst{\ref{aff106}}
\and K.~Paterson\orcid{0000-0001-8340-3486}\inst{\ref{aff66}}
\and L.~Patrizii\inst{\ref{aff4}}
\and A.~Pisani\orcid{0000-0002-6146-4437}\inst{\ref{aff53},\ref{aff150}}
\and V.~Popa\orcid{0000-0002-9118-8330}\inst{\ref{aff151}}
\and D.~Potter\orcid{0000-0002-0757-5195}\inst{\ref{aff152}}
\and P.~Reimberg\orcid{0000-0003-3410-0280}\inst{\ref{aff79}}
\and I.~Risso\orcid{0000-0003-2525-7761}\inst{\ref{aff153}}
\and P.-F.~Rocci\inst{\ref{aff127}}
\and M.~Sahl\'en\orcid{0000-0003-0973-4804}\inst{\ref{aff154}}
\and A.~Schneider\orcid{0000-0001-7055-8104}\inst{\ref{aff152}}
\and M.~Schultheis\inst{\ref{aff76}}
\and D.~Sciotti\orcid{0009-0008-4519-2620}\inst{\ref{aff6},\ref{aff74}}
\and E.~Sellentin\inst{\ref{aff155},\ref{aff156}}
\and M.~Sereno\orcid{0000-0003-0302-0325}\inst{\ref{aff3},\ref{aff4}}
\and A.~Silvestri\orcid{0000-0001-6904-5061}\inst{\ref{aff119}}
\and L.~C.~Smith\orcid{0000-0002-3259-2771}\inst{\ref{aff157}}
\and K.~Tanidis\inst{\ref{aff101}}
\and C.~Tao\orcid{0000-0001-7961-8177}\inst{\ref{aff53}}
\and N.~Tessore\orcid{0000-0002-9696-7931}\inst{\ref{aff68}}
\and G.~Testera\inst{\ref{aff13}}
\and R.~Teyssier\orcid{0000-0001-7689-0933}\inst{\ref{aff150}}
\and S.~Toft\orcid{0000-0003-3631-7176}\inst{\ref{aff158},\ref{aff159}}
\and S.~Tosi\orcid{0000-0002-7275-9193}\inst{\ref{aff15},\ref{aff13}}
\and A.~Troja\orcid{0000-0003-0239-4595}\inst{\ref{aff88},\ref{aff50}}
\and M.~Tucci\inst{\ref{aff48}}
\and C.~Valieri\inst{\ref{aff4}}
\and D.~Vergani\orcid{0000-0003-0898-2216}\inst{\ref{aff3}}
\and G.~Verza\orcid{0000-0002-1886-8348}\inst{\ref{aff160}}
\and P.~Vielzeuf\orcid{0000-0003-2035-9339}\inst{\ref{aff53}}
\and N.~A.~Walton\orcid{0000-0003-3983-8778}\inst{\ref{aff157}}}
										   
\institute{Aix-Marseille Universit\'e, CNRS, CNES, LAM, Marseille, France\label{aff1}
\and
Dipartimento di Fisica e Astronomia "Augusto Righi" - Alma Mater Studiorum Universit\`a di Bologna, via Piero Gobetti 93/2, 40129 Bologna, Italy\label{aff2}
\and
INAF-Osservatorio di Astrofisica e Scienza dello Spazio di Bologna, Via Piero Gobetti 93/3, 40129 Bologna, Italy\label{aff3}
\and
INFN-Sezione di Bologna, Viale Berti Pichat 6/2, 40127 Bologna, Italy\label{aff4}
\and
Department of Physics and Helsinki Institute of Physics, Gustaf H\"allstr\"omin katu 2, 00014 University of Helsinki, Finland\label{aff5}
\and
INAF-Osservatorio Astronomico di Roma, Via Frascati 33, 00078 Monteporzio Catone, Italy\label{aff6}
\and
IFPU, Institute for Fundamental Physics of the Universe, via Beirut 2, 34151 Trieste, Italy\label{aff7}
\and
INAF-Osservatorio Astronomico di Trieste, Via G. B. Tiepolo 11, 34143 Trieste, Italy\label{aff8}
\and
SISSA, International School for Advanced Studies, Via Bonomea 265, 34136 Trieste TS, Italy\label{aff9}
\and
INFN, Sezione di Trieste, Via Valerio 2, 34127 Trieste TS, Italy\label{aff10}
\and
ICSC - Centro Nazionale di Ricerca in High Performance Computing, Big Data e Quantum Computing, Via Magnanelli 2, Bologna, Italy\label{aff11}
\and
INAF-Osservatorio Astronomico di Brera, Via Brera 28, 20122 Milano, Italy\label{aff12}
\and
INFN-Sezione di Genova, Via Dodecaneso 33, 16146, Genova, Italy\label{aff13}
\and
Dipartimento di Fisica, Universit\`a degli studi di Genova, and INFN-Sezione di Genova, via Dodecaneso 33, 16146, Genova, Italy\label{aff14}
\and
Dipartimento di Fisica, Universit\`a di Genova, Via Dodecaneso 33, 16146, Genova, Italy\label{aff15}
\and
Department of Physics, P.O. Box 64, 00014 University of Helsinki, Finland\label{aff16}
\and
Helsinki Institute of Physics, Gustaf H{\"a}llstr{\"o}min katu 2, University of Helsinki, Helsinki, Finland\label{aff17}
\and
INAF-Osservatorio Astronomico di Capodimonte, Via Moiariello 16, 80131 Napoli, Italy\label{aff18}
\and
Institute of Space Sciences (ICE, CSIC), Campus UAB, Carrer de Can Magrans, s/n, 08193 Barcelona, Spain\label{aff19}
\and
School of Mathematics and Physics, University of Surrey, Guildford, Surrey, GU2 7XH, UK\label{aff20}
\and
Dipartimento di Fisica e Astronomia, Universit\`a di Bologna, Via Gobetti 93/2, 40129 Bologna, Italy\label{aff21}
\and
Centre National d'Etudes Spatiales -- Centre spatial de Toulouse, 18 avenue Edouard Belin, 31401 Toulouse Cedex 9, France\label{aff22}
\and
INAF-Osservatorio Astrofisico di Torino, Via Osservatorio 20, 10025 Pino Torinese (TO), Italy\label{aff23}
\and
Department of Physics "E. Pancini", University Federico II, Via Cinthia 6, 80126, Napoli, Italy\label{aff24}
\and
INFN section of Naples, Via Cinthia 6, 80126, Napoli, Italy\label{aff25}
\and
Instituto de Astrof\'isica e Ci\^encias do Espa\c{c}o, Universidade do Porto, CAUP, Rua das Estrelas, PT4150-762 Porto, Portugal\label{aff26}
\and
Faculdade de Ci\^encias da Universidade do Porto, Rua do Campo de Alegre, 4150-007 Porto, Portugal\label{aff27}
\and
Dipartimento di Fisica, Universit\`a degli Studi di Torino, Via P. Giuria 1, 10125 Torino, Italy\label{aff28}
\and
INFN-Sezione di Torino, Via P. Giuria 1, 10125 Torino, Italy\label{aff29}
\and
INAF-IASF Milano, Via Alfonso Corti 12, 20133 Milano, Italy\label{aff30}
\and
Centro de Investigaciones Energ\'eticas, Medioambientales y Tecnol\'ogicas (CIEMAT), Avenida Complutense 40, 28040 Madrid, Spain\label{aff31}
\and
Port d'Informaci\'{o} Cient\'{i}fica, Campus UAB, C. Albareda s/n, 08193 Bellaterra (Barcelona), Spain\label{aff32}
\and
Institute for Theoretical Particle Physics and Cosmology (TTK), RWTH Aachen University, 52056 Aachen, Germany\label{aff33}
\and
Institute of Cosmology and Gravitation, University of Portsmouth, Portsmouth PO1 3FX, UK\label{aff34}
\and
Institut d'Estudis Espacials de Catalunya (IEEC),  Edifici RDIT, Campus UPC, 08860 Castelldefels, Barcelona, Spain\label{aff35}
\and
Dipartimento di Fisica e Astronomia "Augusto Righi" - Alma Mater Studiorum Universit\`a di Bologna, Viale Berti Pichat 6/2, 40127 Bologna, Italy\label{aff36}
\and
Instituto de Astrof\'isica de Canarias, Calle V\'ia L\'actea s/n, 38204, San Crist\'obal de La Laguna, Tenerife, Spain\label{aff37}
\and
Institute for Astronomy, University of Edinburgh, Royal Observatory, Blackford Hill, Edinburgh EH9 3HJ, UK\label{aff38}
\and
Jodrell Bank Centre for Astrophysics, Department of Physics and Astronomy, University of Manchester, Oxford Road, Manchester M13 9PL, UK\label{aff39}
\and
European Space Agency/ESRIN, Largo Galileo Galilei 1, 00044 Frascati, Roma, Italy\label{aff40}
\and
ESAC/ESA, Camino Bajo del Castillo, s/n., Urb. Villafranca del Castillo, 28692 Villanueva de la Ca\~nada, Madrid, Spain\label{aff41}
\and
Universit\'e Claude Bernard Lyon 1, CNRS/IN2P3, IP2I Lyon, UMR 5822, Villeurbanne, F-69100, France\label{aff42}
\and
Institut de Ci\`{e}ncies del Cosmos (ICCUB), Universitat de Barcelona (IEEC-UB), Mart\'{i} i Franqu\`{e}s 1, 08028 Barcelona, Spain\label{aff43}
\and
Instituci\'o Catalana de Recerca i Estudis Avan\c{c}ats (ICREA), Passeig de Llu\'{\i}s Companys 23, 08010 Barcelona, Spain\label{aff44}
\and
UCB Lyon 1, CNRS/IN2P3, IUF, IP2I Lyon, 4 rue Enrico Fermi, 69622 Villeurbanne, France\label{aff45}
\and
Departamento de F\'isica, Faculdade de Ci\^encias, Universidade de Lisboa, Edif\'icio C8, Campo Grande, PT1749-016 Lisboa, Portugal\label{aff46}
\and
Instituto de Astrof\'isica e Ci\^encias do Espa\c{c}o, Faculdade de Ci\^encias, Universidade de Lisboa, Campo Grande, 1749-016 Lisboa, Portugal\label{aff47}
\and
Department of Astronomy, University of Geneva, ch. d'Ecogia 16, 1290 Versoix, Switzerland\label{aff48}
\and
INAF-Istituto di Astrofisica e Planetologia Spaziali, via del Fosso del Cavaliere, 100, 00100 Roma, Italy\label{aff49}
\and
INFN-Padova, Via Marzolo 8, 35131 Padova, Italy\label{aff50}
\and
Universit\'e Paris-Saclay, Universit\'e Paris Cit\'e, CEA, CNRS, AIM, 91191, Gif-sur-Yvette, France\label{aff51}
\and
Space Science Data Center, Italian Space Agency, via del Politecnico snc, 00133 Roma, Italy\label{aff52}
\and
Aix-Marseille Universit\'e, CNRS/IN2P3, CPPM, Marseille, France\label{aff53}
\and
Universit\"ats-Sternwarte M\"unchen, Fakult\"at f\"ur Physik, Ludwig-Maximilians-Universit\"at M\"unchen, Scheinerstrasse 1, 81679 M\"unchen, Germany\label{aff54}
\and
FRACTAL S.L.N.E., calle Tulip\'an 2, Portal 13 1A, 28231, Las Rozas de Madrid, Spain\label{aff55}
\and
INAF-Osservatorio Astronomico di Padova, Via dell'Osservatorio 5, 35122 Padova, Italy\label{aff56}
\and
Max Planck Institute for Extraterrestrial Physics, Giessenbachstr. 1, 85748 Garching, Germany\label{aff57}
\and
Dipartimento di Fisica "Aldo Pontremoli", Universit\`a degli Studi di Milano, Via Celoria 16, 20133 Milano, Italy\label{aff58}
\and
Institute of Theoretical Astrophysics, University of Oslo, P.O. Box 1029 Blindern, 0315 Oslo, Norway\label{aff59}
\and
Jet Propulsion Laboratory, California Institute of Technology, 4800 Oak Grove Drive, Pasadena, CA, 91109, USA\label{aff60}
\and
Felix Hormuth Engineering, Goethestr. 17, 69181 Leimen, Germany\label{aff61}
\and
Technical University of Denmark, Elektrovej 327, 2800 Kgs. Lyngby, Denmark\label{aff62}
\and
Cosmic Dawn Center (DAWN), Denmark\label{aff63}
\and
Universit\'e Paris-Saclay, CNRS/IN2P3, IJCLab, 91405 Orsay, France\label{aff64}
\and
Institut de Recherche en Astrophysique et Plan\'etologie (IRAP), Universit\'e de Toulouse, CNRS, UPS, CNES, 14 Av. Edouard Belin, 31400 Toulouse, France\label{aff65}
\and
Max-Planck-Institut f\"ur Astronomie, K\"onigstuhl 17, 69117 Heidelberg, Germany\label{aff66}
\and
NASA Goddard Space Flight Center, Greenbelt, MD 20771, USA\label{aff67}
\and
Department of Physics and Astronomy, University College London, Gower Street, London WC1E 6BT, UK\label{aff68}
\and
Universit\'e de Gen\`eve, D\'epartement de Physique Th\'eorique and Centre for Astroparticle Physics, 24 quai Ernest-Ansermet, CH-1211 Gen\`eve 4, Switzerland\label{aff69}
\and
NOVA optical infrared instrumentation group at ASTRON, Oude Hoogeveensedijk 4, 7991PD, Dwingeloo, The Netherlands\label{aff70}
\and
Centre de Calcul de l'IN2P3/CNRS, 21 avenue Pierre de Coubertin 69627 Villeurbanne Cedex, France\label{aff71}
\and
INFN-Sezione di Milano, Via Celoria 16, 20133 Milano, Italy\label{aff72}
\and
Universit\"at Bonn, Argelander-Institut f\"ur Astronomie, Auf dem H\"ugel 71, 53121 Bonn, Germany\label{aff73}
\and
INFN-Sezione di Roma, Piazzale Aldo Moro, 2 - c/o Dipartimento di Fisica, Edificio G. Marconi, 00185 Roma, Italy\label{aff74}
\and
Department of Physics, Institute for Computational Cosmology, Durham University, South Road, Durham, DH1 3LE, UK\label{aff75}
\and
Universit\'e C\^{o}te d'Azur, Observatoire de la C\^{o}te d'Azur, CNRS, Laboratoire Lagrange, Bd de l'Observatoire, CS 34229, 06304 Nice cedex 4, France\label{aff76}
\and
Universit\'e Paris Cit\'e, CNRS, Astroparticule et Cosmologie, 75013 Paris, France\label{aff77}
\and
University of Applied Sciences and Arts of Northwestern Switzerland, School of Engineering, 5210 Windisch, Switzerland\label{aff78}
\and
Institut d'Astrophysique de Paris, 98bis Boulevard Arago, 75014, Paris, France\label{aff79}
\and
Institut d'Astrophysique de Paris, UMR 7095, CNRS, and Sorbonne Universit\'e, 98 bis boulevard Arago, 75014 Paris, France\label{aff80}
\and
Institute of Physics, Laboratory of Astrophysics, Ecole Polytechnique F\'ed\'erale de Lausanne (EPFL), Observatoire de Sauverny, 1290 Versoix, Switzerland\label{aff81}
\and
Institut de F\'{i}sica d'Altes Energies (IFAE), The Barcelona Institute of Science and Technology, Campus UAB, 08193 Bellaterra (Barcelona), Spain\label{aff82}
\and
European Space Agency/ESTEC, Keplerlaan 1, 2201 AZ Noordwijk, The Netherlands\label{aff83}
\and
DARK, Niels Bohr Institute, University of Copenhagen, Jagtvej 155, 2200 Copenhagen, Denmark\label{aff84}
\and
Waterloo Centre for Astrophysics, University of Waterloo, Waterloo, Ontario N2L 3G1, Canada\label{aff85}
\and
Department of Physics and Astronomy, University of Waterloo, Waterloo, Ontario N2L 3G1, Canada\label{aff86}
\and
Perimeter Institute for Theoretical Physics, Waterloo, Ontario N2L 2Y5, Canada\label{aff87}
\and
Dipartimento di Fisica e Astronomia "G. Galilei", Universit\`a di Padova, Via Marzolo 8, 35131 Padova, Italy\label{aff88}
\and
Institut f\"ur Theoretische Physik, University of Heidelberg, Philosophenweg 16, 69120 Heidelberg, Germany\label{aff89}
\and
Universit\'e St Joseph; Faculty of Sciences, Beirut, Lebanon\label{aff90}
\and
Departamento de F\'isica, FCFM, Universidad de Chile, Blanco Encalada 2008, Santiago, Chile\label{aff91}
\and
Universit\"at Innsbruck, Institut f\"ur Astro- und Teilchenphysik, Technikerstr. 25/8, 6020 Innsbruck, Austria\label{aff92}
\and
Satlantis, University Science Park, Sede Bld 48940, Leioa-Bilbao, Spain\label{aff93}
\and
Instituto de Astrof\'isica e Ci\^encias do Espa\c{c}o, Faculdade de Ci\^encias, Universidade de Lisboa, Tapada da Ajuda, 1349-018 Lisboa, Portugal\label{aff94}
\and
Universidad Polit\'ecnica de Cartagena, Departamento de Electr\'onica y Tecnolog\'ia de Computadoras,  Plaza del Hospital 1, 30202 Cartagena, Spain\label{aff95}
\and
Centre for Information Technology, University of Groningen, P.O. Box 11044, 9700 CA Groningen, The Netherlands\label{aff96}
\and
INFN-Bologna, Via Irnerio 46, 40126 Bologna, Italy\label{aff97}
\and
Infrared Processing and Analysis Center, California Institute of Technology, Pasadena, CA 91125, USA\label{aff98}
\and
INAF, Istituto di Radioastronomia, Via Piero Gobetti 101, 40129 Bologna, Italy\label{aff99}
\and
Astronomical Observatory of the Autonomous Region of the Aosta Valley (OAVdA), Loc. Lignan 39, I-11020, Nus (Aosta Valley), Italy\label{aff100}
\and
Department of Physics, Oxford University, Keble Road, Oxford OX1 3RH, UK\label{aff101}
\and
Aurora Technology for European Space Agency (ESA), Camino bajo del Castillo, s/n, Urbanizacion Villafranca del Castillo, Villanueva de la Ca\~nada, 28692 Madrid, Spain\label{aff102}
\and
ICL, Junia, Universit\'e Catholique de Lille, LITL, 59000 Lille, France\label{aff103}
\and
Department of Physics, Royal Holloway, University of London, TW20 0EX, UK\label{aff104}
\and
Mullard Space Science Laboratory, University College London, Holmbury St Mary, Dorking, Surrey RH5 6NT, UK\label{aff105}
\and
Instituto de F\'isica Te\'orica UAM-CSIC, Campus de Cantoblanco, 28049 Madrid, Spain\label{aff106}
\and
CERCA/ISO, Department of Physics, Case Western Reserve University, 10900 Euclid Avenue, Cleveland, OH 44106, USA\label{aff107}
\and
Technical University of Munich, TUM School of Natural Sciences, Physics Department, James-Franck-Str.~1, 85748 Garching, Germany\label{aff108}
\and
Max-Planck-Institut f\"ur Astrophysik, Karl-Schwarzschild-Str.~1, 85748 Garching, Germany\label{aff109}
\and
Laboratoire Univers et Th\'eorie, Observatoire de Paris, Universit\'e PSL, Universit\'e Paris Cit\'e, CNRS, 92190 Meudon, France\label{aff110}
\and
Departamento de F{\'\i}sica Fundamental. Universidad de Salamanca. Plaza de la Merced s/n. 37008 Salamanca, Spain\label{aff111}
\and
Departamento de Astrof\'isica, Universidad de La Laguna, 38206, La Laguna, Tenerife, Spain\label{aff112}
\and
Universit\'e de Strasbourg, CNRS, Observatoire astronomique de Strasbourg, UMR 7550, 67000 Strasbourg, France\label{aff113}
\and
Center for Data-Driven Discovery, Kavli IPMU (WPI), UTIAS, The University of Tokyo, Kashiwa, Chiba 277-8583, Japan\label{aff114}
\and
Ludwig-Maximilians-University, Schellingstrasse 4, 80799 Munich, Germany\label{aff115}
\and
Max-Planck-Institut f\"ur Physik, Boltzmannstr. 8, 85748 Garching, Germany\label{aff116}
\and
Dipartimento di Fisica - Sezione di Astronomia, Universit\`a di Trieste, Via Tiepolo 11, 34131 Trieste, Italy\label{aff117}
\and
California institute of Technology, 1200 E California Blvd, Pasadena, CA 91125, USA\label{aff118}
\and
Institute Lorentz, Leiden University, Niels Bohrweg 2, 2333 CA Leiden, The Netherlands\label{aff119}
\and
Institute for Astronomy, University of Hawaii, 2680 Woodlawn Drive, Honolulu, HI 96822, USA\label{aff120}
\and
Department of Physics \& Astronomy, University of California Irvine, Irvine CA 92697, USA\label{aff121}
\and
Department of Mathematics and Physics E. De Giorgi, University of Salento, Via per Arnesano, CP-I93, 73100, Lecce, Italy\label{aff122}
\and
INFN, Sezione di Lecce, Via per Arnesano, CP-193, 73100, Lecce, Italy\label{aff123}
\and
INAF-Sezione di Lecce, c/o Dipartimento Matematica e Fisica, Via per Arnesano, 73100, Lecce, Italy\label{aff124}
\and
Kapteyn Astronomical Institute, University of Groningen, PO Box 800, 9700 AV Groningen, The Netherlands\label{aff125}
\and
Departamento F\'isica Aplicada, Universidad Polit\'ecnica de Cartagena, Campus Muralla del Mar, 30202 Cartagena, Murcia, Spain\label{aff126}
\and
Universit\'e Paris-Saclay, CNRS, Institut d'astrophysique spatiale, 91405, Orsay, France\label{aff127}
\and
CEA Saclay, DFR/IRFU, Service d'Astrophysique, Bat. 709, 91191 Gif-sur-Yvette, France\label{aff128}
\and
Department of Computer Science, Aalto University, PO Box 15400, Espoo, FI-00 076, Finland\label{aff129}
\and
Instituto de Astrof\'\i sica de Canarias, c/ Via Lactea s/n, La Laguna E-38200, Spain. Departamento de Astrof\'\i sica de la Universidad de La Laguna, Avda. Francisco Sanchez, La Laguna, E-38200, Spain\label{aff130}
\and
Caltech/IPAC, 1200 E. California Blvd., Pasadena, CA 91125, USA\label{aff131}
\and
Ruhr University Bochum, Faculty of Physics and Astronomy, Astronomical Institute (AIRUB), German Centre for Cosmological Lensing (GCCL), 44780 Bochum, Germany\label{aff132}
\and
Univ. Grenoble Alpes, CNRS, Grenoble INP, LPSC-IN2P3, 53, Avenue des Martyrs, 38000, Grenoble, France\label{aff133}
\and
Department of Physics and Astronomy, Vesilinnantie 5, 20014 University of Turku, Finland\label{aff134}
\and
Serco for European Space Agency (ESA), Camino bajo del Castillo, s/n, Urbanizacion Villafranca del Castillo, Villanueva de la Ca\~nada, 28692 Madrid, Spain\label{aff135}
\and
ARC Centre of Excellence for Dark Matter Particle Physics, Melbourne, Australia\label{aff136}
\and
Centre for Astrophysics \& Supercomputing, Swinburne University of Technology,  Hawthorn, Victoria 3122, Australia\label{aff137}
\and
Dipartimento di Fisica e Scienze della Terra, Universit\`a degli Studi di Ferrara, Via Giuseppe Saragat 1, 44122 Ferrara, Italy\label{aff138}
\and
Department of Physics and Astronomy, University of the Western Cape, Bellville, Cape Town, 7535, South Africa\label{aff139}
\and
Istituto Nazionale di Fisica Nucleare, Sezione di Ferrara, Via Giuseppe Saragat 1, 44122 Ferrara, Italy\label{aff140}
\and
DAMTP, Centre for Mathematical Sciences, Wilberforce Road, Cambridge CB3 0WA, UK\label{aff141}
\and
Kavli Institute for Cosmology Cambridge, Madingley Road, Cambridge, CB3 0HA, UK\label{aff142}
\and
IRFU, CEA, Universit\'e Paris-Saclay 91191 Gif-sur-Yvette Cedex, France\label{aff143}
\and
Oskar Klein Centre for Cosmoparticle Physics, Department of Physics, Stockholm University, Stockholm, SE-106 91, Sweden\label{aff144}
\and
Astrophysics Group, Blackett Laboratory, Imperial College London, London SW7 2AZ, UK\label{aff145}
\and
INAF-Osservatorio Astrofisico di Arcetri, Largo E. Fermi 5, 50125, Firenze, Italy\label{aff146}
\and
Dipartimento di Fisica, Sapienza Universit\`a di Roma, Piazzale Aldo Moro 2, 00185 Roma, Italy\label{aff147}
\and
Centro de Astrof\'{\i}sica da Universidade do Porto, Rua das Estrelas, 4150-762 Porto, Portugal\label{aff148}
\and
HE Space for European Space Agency (ESA), Camino bajo del Castillo, s/n, Urbanizacion Villafranca del Castillo, Villanueva de la Ca\~nada, 28692 Madrid, Spain\label{aff149}
\and
Department of Astrophysical Sciences, Peyton Hall, Princeton University, Princeton, NJ 08544, USA\label{aff150}
\and
Institute of Space Science, Str. Atomistilor, nr. 409 M\u{a}gurele, Ilfov, 077125, Romania\label{aff151}
\and
Department of Astrophysics, University of Zurich, Winterthurerstrasse 190, 8057 Zurich, Switzerland\label{aff152}
\and
INAF-Osservatorio Astronomico di Brera, Via Brera 28, 20122 Milano, Italy, and INFN-Sezione di Genova, Via Dodecaneso 33, 16146, Genova, Italy\label{aff153}
\and
Theoretical astrophysics, Department of Physics and Astronomy, Uppsala University, Box 515, 751 20 Uppsala, Sweden\label{aff154}
\and
Mathematical Institute, University of Leiden, Niels Bohrweg 1, 2333 CA Leiden, The Netherlands\label{aff155}
\and
Leiden Observatory, Leiden University, Einsteinweg 55, 2333 CC Leiden, The Netherlands\label{aff156}
\and
Institute of Astronomy, University of Cambridge, Madingley Road, Cambridge CB3 0HA, UK\label{aff157}
\and
Cosmic Dawn Center (DAWN)\label{aff158}
\and
Niels Bohr Institute, University of Copenhagen, Jagtvej 128, 2200 Copenhagen, Denmark\label{aff159}
\and
Center for Computational Astrophysics, Flatiron Institute, 162 5th Avenue, 10010, New York, NY, USA\label{aff160}}  

\date{Received XXX / Accepted YYY}

\abstract{
The 2-point correlation function of the galaxy spatial distribution is a major cosmological observable that enables constraints on the dynamics and geometry of the Universe. The \Euclid mission aims at performing an extensive spectroscopic survey of approximately 20--30 million H$\alpha$-emitting galaxies up to about redshift two. This ambitious project seeks to elucidate the nature of dark energy by mapping the 3-dimensional clustering of galaxies over a significant portion of the sky. This paper presents the methodology and software developed for estimating the 3-dimensional 2-point correlation function within the Euclid Science Ground Segment. The software is designed to overcome the significant challenges posed by the large and complex \Euclid data set, which involves millions of galaxies. Key challenges include efficient pair counting, managing computational resources, and ensuring the accuracy of the correlation function estimation. The software leverages advanced algorithms, including kd-tree, octree, and linked-list data partitioning strategies, to optimise the pair-counting process. These methods are crucial for handling the massive volume of data efficiently. The implementation also includes parallel processing capabilities using shared-memory open multi-processing to further enhance performance and reduce computation times. Extensive validation and performance testing of the software are presented. Those have been performed by using various mock galaxy catalogues to ensure that it meets the stringent accuracy requirement of the \Euclid mission. The results indicate that the software is robust and can reliably estimate the 2-point correlation function, which is essential for deriving cosmological parameters with high precision. Furthermore, the paper discusses the expected performance of the software during different stages of the Euclid Wide Survey observations and forecasts how the precision of the correlation function measurements will improve over the mission's timeline, highlighting the software's capability to handle large data sets efficiently.}


\titlerunning{\Euclid preparation. 2-point correlation function estimation}

\authorrunning{de la Torre at al.}

\maketitle

\section{Introduction}

Galaxies in the Universe tend to cluster as they evolve within the large-scale structure, itself growing under the influence of gravity and universal expansion. Their observed positions are possibly the best tracers of the overall 3-dimensional matter spatial distribution, given their high and unrivalled observed number density, among all observable objects on cosmological scales. This makes galaxy clustering and its features crucial for studying the expansion and structure growth in the late Universe \citep[e.g.,][]{Amendola18}. In particular, the baryon acoustic oscillations (BAO) feature used as a standard ruler, makes galaxy clustering sensitive to the geometry and energy content of the Universe \citep[e.g.,][]{alam21}. Moreover, peculiar velocities contribute to observed galaxy redshifts and distort the genuine spatial distribution along the line-of-sight direction. On large scales, these velocities map the coherent flows towards over-densities induced by the growth of structure, whose strength is dictated by the laws of gravity. This apparent feature is of paramount importance in testing standard gravity and the cosmological model \citep[][]{kaiser87,guzzo08}.

Galaxy clustering can be quantified in a statistical manner by determining the $n$-point statistics of the spatial distribution of galaxies. This is achieved in configuration space by estimating the 2-point correlation function (2PCF), as well as higher-order $n$-point correlation functions, which enable the exploration of non-Gaussian features of the galaxy spatial distribution. The 3-dimensional 2PCF is, with the power spectrum, the most commonly used tool for analysing galaxy spectroscopic survey data and inferring cosmological model parameters. For this, it is crucial to estimate the  contributions of the 2PCF along the parallel and transverse directions to the line-of-sight, as the apparent spatial distribution of galaxies is anisotropic.

The \Euclid mission is a medium-sized European Space Agency (ESA) space mission to unravel the dark sector of the Universe \citep{EuclidSkyOverview}. It is primarily devoted to the investigation of the nature of dark energy and the dark matter distribution. The \Euclid satellite was launched on July 1st of 2023 and will survey about $\num{14\,000}$ deg$^2$ of the extragalactic sky, performing one of the largest galaxy surveys ever made. It will probe the last 10 billion years of the universal expansion history via its main cosmological probes: weak gravitational lensing and galaxy clustering. In particular, \Euclid's Near-Infrared Spectrometer and Photometer \citep[NISP,][]{EuclidSkyNISP} will measure the spectroscopic redshifts of about 20--30 million H$\alpha$-emitting galaxies using grism spectroscopy. The resulting 3-dimensional galaxy map will serve to characterise the clustering of matter, and in turn determine cosmological parameters with an unprecedented precision. In this, the estimation of the 2PCF of spectroscopic galaxies will play a crucial role. Nonetheless, the estimation of the 2PCF from such a large data set presents a number of challenges, particularly in terms of the efficiency of the estimation. In order to address those challenges, we have designed a robust, highly-tested, and efficient software that can be used within the Euclid Science Ground Segment (SGS). This pipeline element, which provides one of the critical end-products for \Euclid scientific exploitation, is the processing function 2PCF-GC. 

This paper presents the method and 2PCF-GC processing function software devised to estimate the 3-dimensional 2PCF within the Euclid SGS pipeline. It describes the ensemble of tests that have been performed to ensure the highest accuracy and efficiency in the estimation. Furthermore, it provides a discussion on how the 2PCF-GC processing function addresses \Euclid computational challenges and provides forecasts on real survey 2PCF estimation. In the following, we will simply refer to the 2PCF-GC processing function and software as 2PCF-GC. This paper is part of a set that describes all \Euclid galaxy clustering processing functions, which also includes those for three-point correlation function (Veropalumbo et al., in prep.), as well as power spectrum and bispectrum (Salvalaggio et al., in prep.) estimations.

The paper is structured as follows. Section 2 presents the 2PCF definition and estimators. Section 3 discusses the optimisation of the software. Section 4 describes the validation and performance tests as well as associated results. Section 5 shows predictions of 2PCF measurements in the Euclid Wide Survey, and finally Sect. 6 provides conclusions.

\section{2-point correlation function}

\subsection{Definition and estimators}

The 2PCF is defined as the probability of finding two objects at volume elements $\diff V_1$ and $\diff V_2$ separated by a vector $\br$, with respect to a Poisson random distribution. This excess probability is defined as \citep{peebles80}
\begin{equation}
   \diff P = n^2 \left[ 1+\xirv \right] \diff V_1 \diff V_2 \, , 
\end{equation}
where $n$ is the object mean number density. $\xirv$ therefore measures the clustering in excess or in deficit compared to a random Poisson point distribution in space. In the previous equation, statistical homogeneity is implicitly assumed in that the 2PCF is only a function of the separation vector, and not on the positions of the two objects.

In 2PCF-GC, we use the minimum-variance estimator from \citet{landy93},
which allows the estimation of the 2-point auto-correlation function for a set of spatially distributed objects, based on pair counts as a function of separation. As opposed to other proposed estimators, it minimises the estimator variance and mitigates discreteness and edge effects. This estimator is defined as
\begin{equation}
   \xirv = \frac{\mathrm{DD}(\br) - 2 \, \mathrm{DR}(\br) + \mathrm{RR}(\br)}{\mathrm{RR}(\br)}\, ,
\end{equation}
where $\mathrm{DD}$, $\mathrm{DR}$, and $\mathrm{RR}$ correspond to data-data, data-random, and random-random normalised (distinct) pair counts, respectively. It requires a catalogue of unclustered objects (random catalogue hereafter) that randomly samples the effective volume of the data catalogue, for which we want to estimate the 2PCF. Schematically, the \citet{landy93} estimator can be derived by taking the 2-point correlation of the catalogue density contrast $\delta$ defined from data and random catalogues, such that $\delta = (D - R)/R$ where $D$ and $R$ represent respectively the data and associated random catalogue counts. The raw pair counts need to be normalised to the total number of pairs. Let $N_{\rm D}$ and $N_{\rm R}$ be the number of objects in the data and random catalogues respectively, the normalised counts are obtained as
\begin{align}
   \mathrm{DD}(\br) &= \frac{2 \, \widehat{\mathrm{DD}} (\br)}{N_{\rm D}(N_{\rm D}-1)}\, , \\
   \mathrm{DR}(\br) &= \frac{\widehat{\mathrm{DR}}(\br)}{N_{\rm D} N_{\rm R}}\, , \\
   \mathrm{RR}(\br) &= \frac{2 \, \widehat{\mathrm{RR}}(\br)}{N_{\rm R}(N_{\rm R}-1)}\, ,
\end{align}
where $\widehat{\mathrm{DD}}$, $\widehat{\mathrm{DR}}$, $\widehat{\mathrm{RR}}$ correspond to raw counts. These factors result from the fact that the number of distinct pairs that can be drawn from $N$ objects is $N(N-1)/2$ and the number of cross pairs that can be drawn from two sets of $N_1$ and $N_2$ objects is $N_1 N_2$. In the presence of object weights $w$, such as when using \citet{FKP} weights, each pair in the counts contribute with the multiplication of individual object weights and the normalised counts become:
\begin{align}
   \mathrm{DD}(\br) &= \frac{2 \, \widehat{\mathrm{DD}}(\br)}{\left(\sum_{i=1}^{N_{\rm D}} w_i\right)^2 -  \sum_{i=1}^{N_{\rm D}} w^2_i}\, , \\
   \mathrm{DR}(\br) &= \frac{\widehat{\mathrm{DR}}(\br)}{\left(\sum_{i=1}^{N_{\rm D}} w_i \right) \left(\sum_{i=1}^{N_{\rm R}} w_i\right)}\, , \\
   \mathrm{RR}(\br) &= \frac{2 \, \widehat{\mathrm{RR}}(\br)}{\left(\sum_{i=1}^{N_{\rm R}} w_i\right)^2 -  \sum_{i=1}^{N_{\rm R}} w^2_i}\, ,
\end{align}
where the index $i$ goes over the objects in the catalogues.

One may also want to measure the cross-correlation function between two sets of objects in an overlapping volume. For this, one can devise a cross-correlation estimator in a similar fashion as for the auto-correlation case. By defining the density contrasts for the two populations $\delta_1 = (D_1 - R_1)/R_1$ and $\delta_2 = (D_2 - R_2)/R_2$, where now $D_1$ ($D_2$) and $R_1$ ($R_2$) stand for the data and random catalogue counts of the population $1$ ($2$), the 2-point cross-correlation function estimator reads
\begin{equation}
    \xi_{12}(\br) = \frac{\mathrm{D}_1 \mathrm{D}_2 (\br) - \mathrm{D}_1 \mathrm{R}_2 (\br) - \mathrm{R}_1 \mathrm{D}_2 (\br) + \mathrm{R}_1 \mathrm{R}_2(\br)}{\mathrm{R}_1 \mathrm{R}_2(\br)}\, ,
\end{equation}
where $\mathrm{D}_1 \mathrm{D}_2$, $\mathrm{D}_1 \mathrm{R}_2$, $\mathrm{R}_1 \mathrm{D}_2$, and $\mathrm{R}_1 \mathrm{R}_2$ are the data 1-data 2, data 1-random 2, random 1-data 2 and random 1-random 2 normalised pair counts, respectively.

For some purposes, it can also be useful to define a modified density contrast that uses two different random catalogues. Such a modification allows the subtraction of known correlations from the data catalogue, for instance for the purpose of building BAO reconstruction estimators \citep{padmanabhan12} or correcting for observational systematic errors \citep[e.g.,][]{paviot22}. If we thus define an auxiliary random catalogue $S$ and modified density contrast $\delta^{\rm m} = (D - S)/R$, the estimator for the auto-correlation function associated to that density contrast becomes
\begin{equation}
    \xi^{\rm m}(\br) = \frac{\mathrm{DD} (\br) - 2 \, \mathrm{DS} (\br) + \mathrm{SS}(\br)}{\mathrm{RR}(\br)}\, .
\end{equation}
Similarly, the cross-correlation function estimator associated with the modified density contrast reads
\begin{equation}
    \xi^{\rm m}_{12}(\br) = \frac{\mathrm{D}_1 \mathrm{D}_2 (\br) - \mathrm{D}_1 \mathrm{S}_2 (\br) - \mathrm{S}_1 \mathrm{D}_2 (\br) + \mathrm{S}_1 \mathrm{S}_2(\br)}{\mathrm{R}_1 \mathrm{R}_2(\br)}\, .
\end{equation}

The pair counts and estimated 2PCF with these estimators involve a binning scheme in the pair separation. In 2PCF-GC, the latter can be set to either linear or logarithmic base $10$. For instance, in the case of the angle-averaged\footnote{This is the 2PCF as a function of the norm of the separation vector.} 2PCF with linear binning in $r$, the notation $\xir$ will refer to the 2PCF for separations between $r - \Delta r / 2$ and $r + \Delta r / 2$, where $\Delta r$ is the constant linear bin size. In the case of logarithmic binning, it will refer to the 2PCF for logarithms of the separation between $\logten (r/\mathrm{u}) - \Delta \logten (r/\mathrm{u}) / 2$ and $\logten (r/\mathrm{u}) + \Delta \logten (r/\mathrm{u}) / 2$, where $\Delta \logten (r/\mathrm{u})$ is the constant logarithmic bin size and $\mathrm{u}$ is the fiducial unit of length.

Finally, it is important to emphasise that all previous estimators are exact and unbiased in the limit of infinite random catalogues. From the computational perspective, those estimators reduce to counting pairs as a function of separation from different catalogues. It is clear from the definitions that the main technical challenge in estimating the 2PCF is the ability of counting all pairs as a function of separation from large catalogues. Naively, this process would scale as $N^2$, where $N$ is the number of objects in the catalogue. This can become intractable in the case of very large catalogues, typically with numbers of object above $10^6$, but efficient algorithms can reduce the amplitude of the scaling.

\subsection{Scale dependence}

The 2PCF is a function of the separation vector $\br$ and under the assumption of an isotropic universe, the 2PCF may only depend on the norm of the separation vector $r=|\br|$. Nonetheless, correlations are affected by apparent physical anisotropies, such as redshift-space distortions, which we want to quantify. For this, the 3-dimensional separation vector can be decomposed into transverse and parallel to the line-of-sight directions. We define the transverse and parallel separation vectors $\Vec{r}_\perp$ and $\Vec{r}_\parallel$ such that
\begin{equation} \label{eq:rdecomp1}
\br = \Vec{r}_\perp + \Vec{r}_\parallel
\end{equation}
and 
\begin{align}
    \Vec{r}_\parallel &\equiv \rpa \, \Vec{u}_\parallel = (\br \cdot \Vec{u}_\parallel) \, \Vec{u}_\parallel\, , \\
    \Vec{r}_\perp &\equiv \rpe \, \Vec{u}_\perp = \Vec{r} - (\br \cdot \Vec{u}_\parallel) \, \Vec{u}_\parallel\, ,
\end{align}
where $\br = \Vec{x}_2-\Vec{x}_1$, $\Vec{x}_1$ and $\Vec{x}_2$ are the 3-dimensional positions of the two objects, $\bx$ is the pair line-of-sight vector (defined below in Eq. \ref{eq:losdef}), $\Vec{u}_\parallel = \bx/|\bx|$, and $\Vec{u}_\perp$ is a unit vector perpendicular to $\Vec{u}_\parallel$ and belonging to the plane determined by the non-collinear vectors $\br$ and $\bx$. With these definitions we can define the anisotropic 2PCF $\xirppi$.

\begin{figure}[htbp!]
\centering
\begin{minipage}[b]{.3\hsize}
    \centering
    \includegraphics[angle=0,width=\hsize]{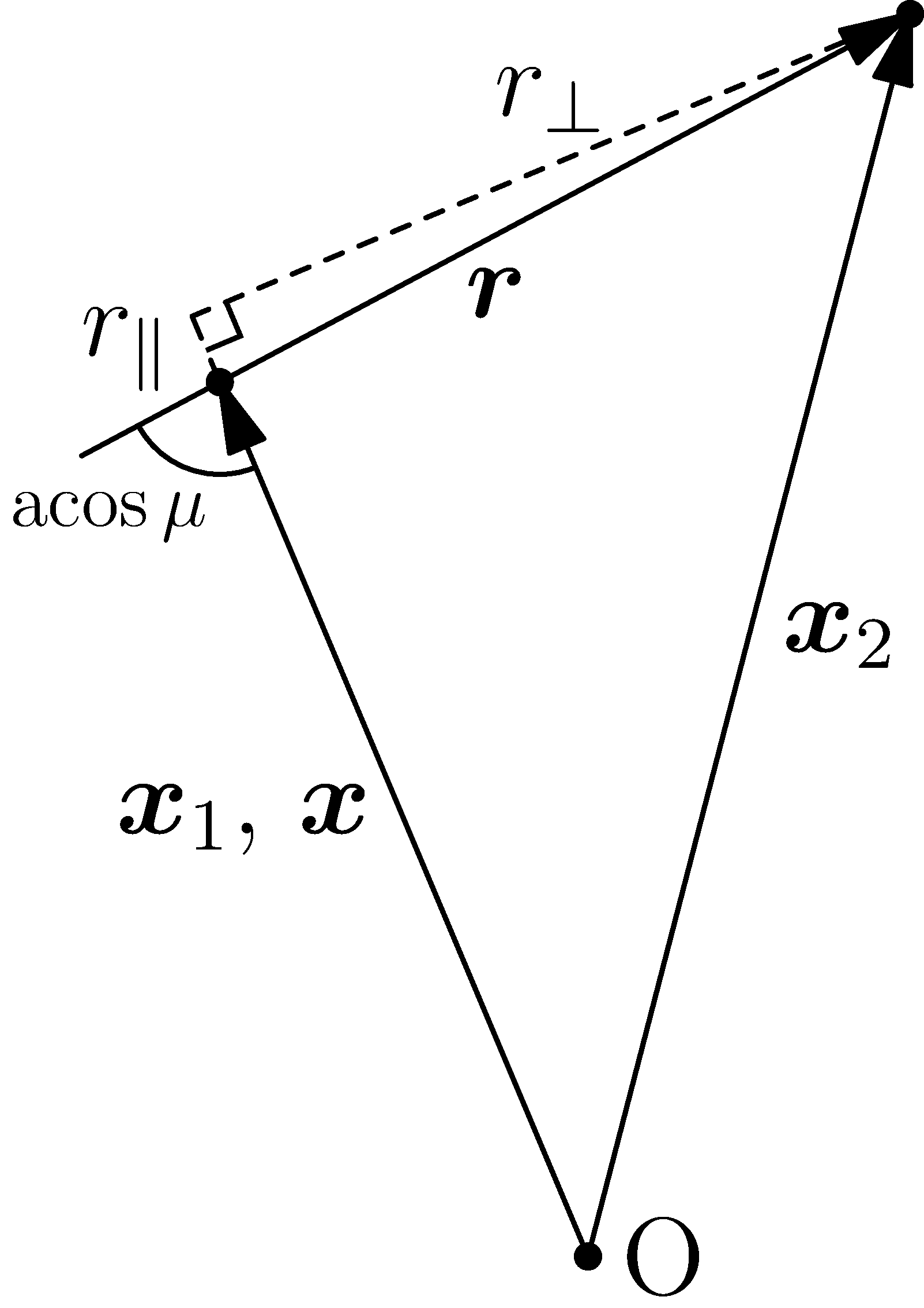} \\
    end-point
\end{minipage}
\begin{minipage}[b]{.3\hsize}
    \centering
    \includegraphics[angle=0,width=\hsize]{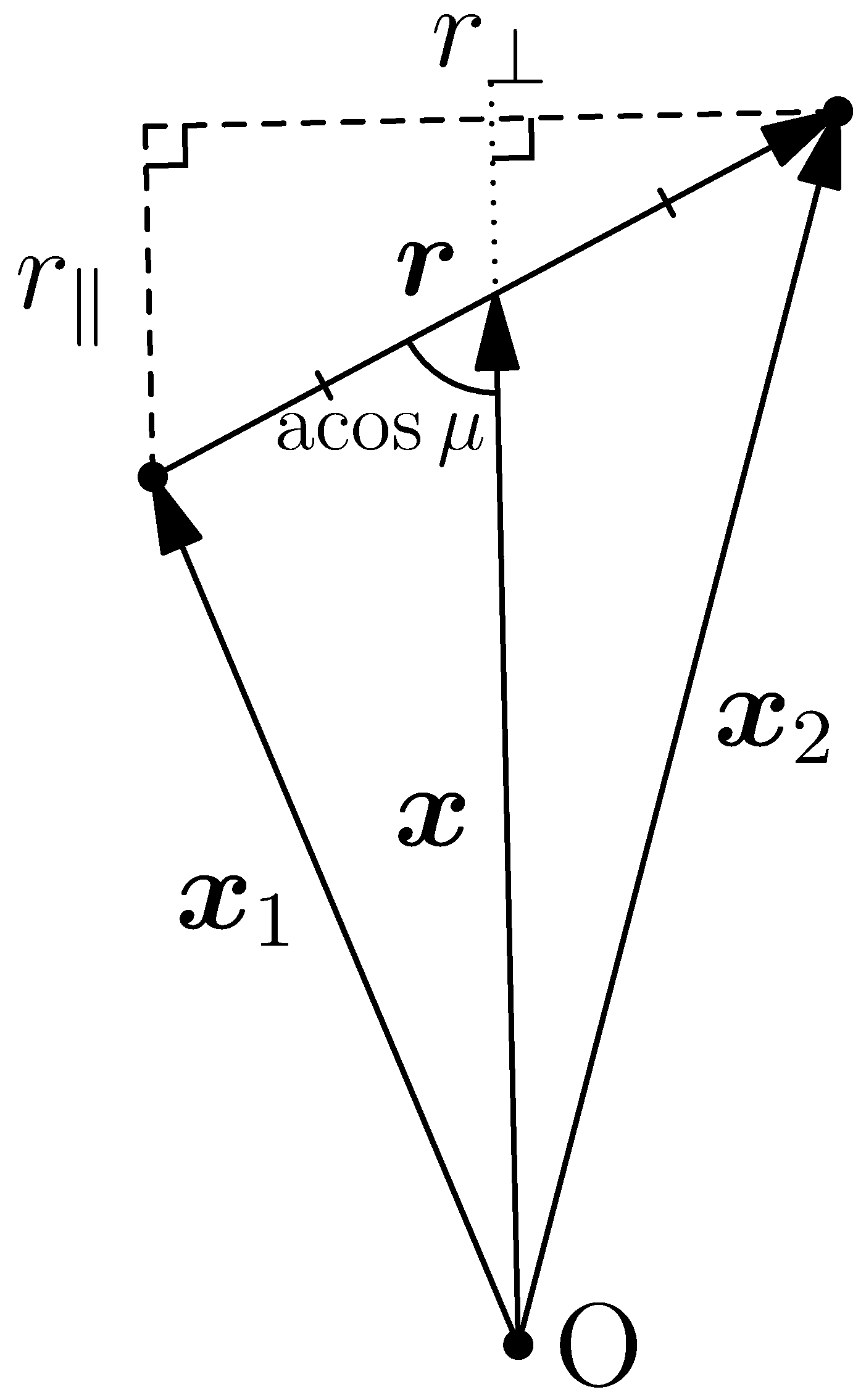} \\
    mid-point
\end{minipage}
\begin{minipage}[b]{.3\hsize}
    \centering
    \includegraphics[angle=0,width=\hsize]{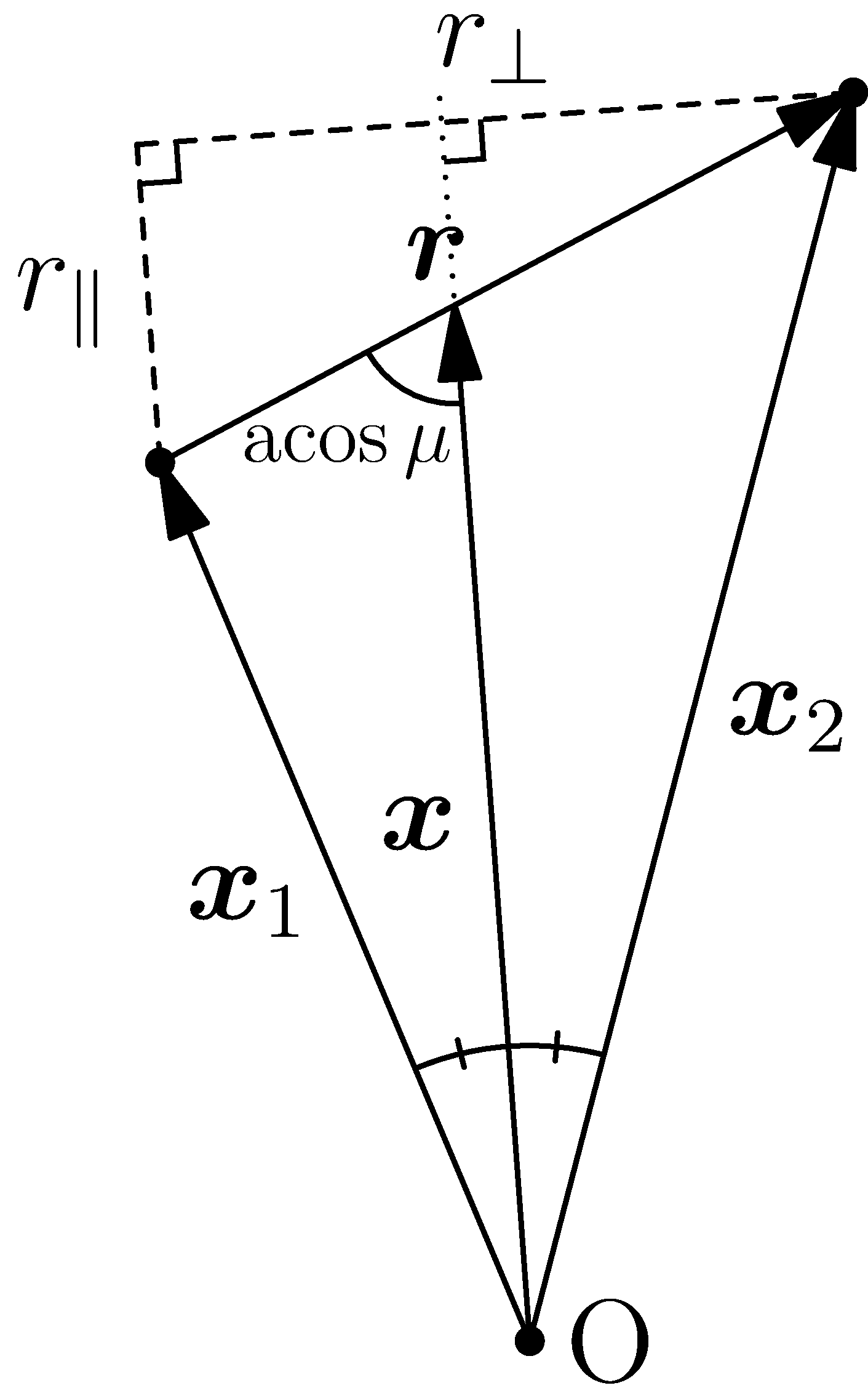} \\
    bisector
\end{minipage}
\caption{Geometry and separation definition for end-point (left), mid-point (middle), and bisector (right) pair line-of-sight conventions. The observer is at the point 'O' and the two other points represent the objects of the pair. In those figures, all vectors and segments are contained in the plane determined by the vectors $\Vec{x}$ and $\Vec{r}$.}
\label{fig:geometry}
\end{figure}

In the decomposition presented in Eq. \eqref{eq:rdecomp1}, we made the implicit assumption that all galaxy lines of sight are parallel, and we defined a mean pair line-of-sight direction $\bx$. This choice is motivated in the distant-observer limit, when $\rpe \ll \rpa$. In full generality, one should consider the dependence on the triangular configuration comprising the observer and the two objects of the pair. This would require the use of three scalars to define the pair configuration. The systematic error in the correlation function estimated within the distant-observer approximation, usually referred to as wide-angle effects is, however, only relevant at large transverse separations and is generally neglected. Usual prescriptions\footnote{Alternatively, one can define $\rpa = |\Vec{x}_2| - |\Vec{x}_1|$, where in that case, there is no explicit need to define a pair line-of-sight vector.} for the pair line-of-sight definition are the mid-point, bisector, or end-point definitions, the mid-point being the most commonly used. These correspond to using
\begin{equation} \label{eq:losdef}
  \bx =
\begin{cases}
    \Vec{x}_1/2 +\Vec{x}_2/2 & \text{for mid-point}\, , \\
    \Vec{x}_1  & \text{for end-point}\, ,\\
    \Vec{x}_1/|\Vec{x}_1|+\Vec{x}_2/|\Vec{x}_2| & \text{for bisector}\, .
\end{cases}
\end{equation}
The geometry and different definitions are made explicit in Fig. \ref{fig:geometry}. Given the symmetries present in the mid-point and bisector definitions, they are less sensitive to wide-angle distortions than the end-point one. Those three definitions use the so-called local plane-parallel approximation, while using a single constant line-of-sight direction for all pairs in a catalogue would lead to the so-called global plane-parallel approximation \citep{samushia15}. The standard approach used in 2PCF-GC is to use the mid-point definition but the other definitions can be chosen for specific cases.

Instead of directly using a decomposition of the separation vector around the line-of-sight direction, it can be useful to express the separation vector in the associated polar basis with coordinates
\begin{align}
    r &= |\br| \\
    \mu &= \rpa/|\br|\, ,   
\end{align}
where $\mu$ is the cosine angle between the separation and line-of-sight directions (see Fig. \ref{fig:geometry}). From this we can further define the 2PCF multipole moments of order $\ell$
\begin{equation} \label{eq:xiell}
    \xi_\ell(r)=\frac{2\ell+1}{2} \int_{-1}^{1}  \diff\mu \, \xismu \, L_\ell(\mu)\, ,
\end{equation}
where $L_\ell(\mu)$ is the Legendre polynomial of order $\ell$. In the auto-correlation function case, there is no physically expected odd multipole signal. However, in the case of the cross-correlation between different populations, a relativistic signal can arise in the odd multipole correlation functions \citep[e.g.,][]{breton19}. 2PCF-GC has been built such that this type of signal can be extracted and provides in output all even and odd moments up to $\ell=4$.

Finally, an estimate of real-space clustering, i.e. without the effect of redshift-space distortions, can be obtained by integrating $\xirppi$ over $\rpa$. This leads to the projected 2PCF
\begin{equation} \label{eq:wprp}
    \wprp = \int_{-\infty}^{\infty} \diff \rpa \, \xirppi\, , 
\end{equation}
where in practice the integral is definite and finite integration limits have to be defined to mitigate the effect of large uncorrelated pairs in the integral. In 2PCF-GC, both integrals in Eqs. \eqref{eq:xiell} and \eqref{eq:wprp} are evaluated as Riemman sums over the linearly binned anisotropic 2PCF.

\subsection{Data spatial partitioning}

The implementation of the 3-dimensional 2PCF estimation in \Euclid uses specific data partitionings to enable an efficient estimation from the huge data set that \Euclid will produce. An overall spectroscopic sample of about 20--30 million galaxies with redshifts is expected in the completed Euclid Wide Survey \citep{EuclidSkyOverview}. To address this task three efficient pair-counting algorithms have been developed, based on different data spatial partitionings: \emph{linked-list}, \emph{kd-tree}, and \emph{octree}. They exploit the observation that, at fixed requested scale range, not all possible distinct pairs have to be computed and stored. Those data spatial partitionings allow us to explore efficiently all pairs where the separation falls within the requested scale range and prune the exploration of irrelevant data. The purpose of developing different pair-counting methods, is to have different ways of assessing the same quantity and identify the fastest and more reliable method. After the development and optimisation of the methods, we found that in the end all three are very efficient, as discussed in the next section. It is worth emphasising that those methods are exact and no approximation is involved at the pair-counting level.

\subsubsection{Linked-list algorithm}

The linked-list algorithm for range searching, also sometimes referred to as the chained-mesh algorithm, implements a 3-dimensional regular pixellation scheme. Our implementation builds on the work of \citet{marulli16} and a similar algorithm has been used in other implementations \citep[e.g.,][]{alonso12,sinha20}. The data bounding volume is divided into a regular Cartesian mesh, and the indexes of objects residing in each cell are stored in a list: the linked list. In practice, the elements (or nodes) of the list are not explicitly linked, instead they are stored in a vector of indexes that map to their 3-dimensional positions.

The pair counting is performed through two nested loops: a first loop goes through all cells containing objects, while the second explores the neighbouring cells whose maximum distance is below the requested maximum separation. In the inner loop, all pairs of objects are counted as in the naive nested-loop algorithm. This strategy allows the pruning of irrelevant pair counts that are outside of the required separation range. The efficiency of the algorithm depends on the cell size and mean number of object per cell. Optimally, one would like to have a cell volume corresponding to a multiple of the search sphere and the most appropriate mean number of object per cell. The latter has to be large enough to avoid having to explore too many cell-cell pairs but small enough to avoid having a too large cell volume, and in turn the pruning to be inefficient. After several trials, an optimal choice of setting the number of cells such that there are about $100$ objects per cell on average is used in our implementation. Practically, this is done by first taking the maximum separation as cell size and estimating the averaged number of objects per cell. Then, the cell size is divided by the integer value that allows us to reach approximately $100$ objects per cell on average.

\subsubsection{Kd-tree algorithm}

The kd-tree range-search algorithm \citep{bentley75} is based on partitioning the set of object positions into a binary tree. Starting from the smallest axis-aligned bounding volume encompassing all data points, i.e. the tree root, tree nodes are obtained by recursively dividing into two equipopulated subsets until reaching the leaves of the tree. Each subsequent partitioning forms two children nodes that contain subsets of the parent node set, and in which objects are close in space. Each binary split increases by unity the depth of the tree. In the classical kd-tree, the node splitting is performed along the dimension with largest spread, at the median object position. The leaves correspond to the highest-depth nodes, when the number of objects reaches a minimum value. In our implementation we use a minimum value of $100$ objects and the sliding-midpoint method for the node splitting rule, which is more adapted for clustered objects \citep{maneewongvatana99}. This choice has proven to be the most efficient.

A crucial aspect of kd-tree is that the bounding volume coordinates of each node are kept in the data structure. This information is then used to search through the tree. For the purpose of counting pairs, we use the dual-tree approach that is a generalisation of the single-tree range-search algorithm for pairs \citep{moore00}. The search is performed by spanning two trees simultaneously and testing each node pair recursively. Depending of the minimum and maximum distance between the tested nodes, the search can be stopped or passed through to the children nodes. This process stops when reaching the leaves. With this method, one can efficiently prune pairs, when separations are outside of the required separation range. The pair counting is effectively performed at the level of the leaves, using a nested loop going over all pairs, whose complexity goes as $\mathcal{O}(n^2)$, where $n$ is the number of objects in the leaves.

\subsubsection{Octree algorithm}

In the octree algorithm \citep{meagher80}, the partitioning of the data volume is organised in a tree structure in which each internal node has exactly eight children, as opposed to two children for the kd-tree. The octree is built from the smallest cubic bounding volume encompassing all the data points. This root node is then recursively subdivided in octants, obtained by dividing the side in each of the three dimensions in two equal parts. Each subsequent partitioning increases the depth of the tree and this process stops when reaching the leave nodes. The leaves can be defined by imposing the maximum depth of the tree, or a minimum value for the number objects in the leaves. Similarly as for the kd-tree, and after several trials, we set the latter minimum value to $\num{100}$ objects.

Our implementation uses a hashed octree structure \citep{warren93}, where tree nodes are stored in a hash table and each node is identified by its Morton binary code \citep{morton66}. This allows the optimisation of the data structure storage and memory access, as well as leads to best performance in tree traversing. Because of the regularity of the octree spatial structure, the node bounding volume coordinates can efficiently be deduced from the Morton code by using bitwise operations. This information is then used to search through the tree. The pair counting is performed similarly as for the kd-tree, by spanning two octrees simultaneously and testing each node pair recursively. Depending of the minimum and maximum distance between the tested nodes, the search can be stopped or passed through to the children nodes, and the process stops when reaching the leaves. The pair counting is performed at the level of leaves, in a similar manner as for the kd-tree.

\subsection{Software architecture}

The 2PCF-GC processing function is a processing element of the \Euclid SGS, which carries out the entire data processing up to cosmological parameter extraction. There are ten Organisational Units (OU) within the SGS, each one having the responsibility to define, design, and validate a specific analysis of the SGS workflow. 2PCF-GC belongs to OU-Level 3 that is in charge of producing the highest-level scientific data products. The data processing within the \Euclid SGS is performed in a distributed system across the  Science Data Centers (SDC) from Finland, France, Germany, Italy, Netherlands, Spain, Switzerland, United Kingdom, and the United States. While most of the SDC have a High Throughput Computing (HTC) design, the underlying infrastructure can vary across SDC. For high-performance computation, as particularly required by this processing function, the overall infrastructure allows for parallelisation.

The development of the 2PCF-GC code was performed in C++11 within the framework of the \Euclid SGS common tools and guidelines to ensure homogeneity in terms of development, storage, and computing, independently of the location \citep{frailis19}. This development followed the common Euclid Development ENvironment (EDEN), which establishes the set of libraries and associated versions to be used by any of the \Euclid software and prevents inconsistencies or changes in the functionality of different libraries between development and production. 2PCF-GC was integrated in the COllaborative DEvelopment ENvironment (CODEEN), a continuous integration and delivery (CI/CD) platform that automates the building, unit testing, and distribution of all the scientific software in the SGS. The source code is stored in a Version Control System (Gitlab) and can be run through a CI/CD pipeline to be finally deployed on a distributed file system available on all SDCs. This system design allows SGS operations to be performed smoothly and efficiently across all SDC, providing the extra advantage of increased computing power and storage capacity. 2PCF-GC has been developed to run within the SGS infrastructure, but can also be run in standalone mode, which has demonstrated to be crucial for testing and validation activities.

\subsubsection{Process overview}

\begin{figure*}
\centering
\includegraphics[width=\textwidth] {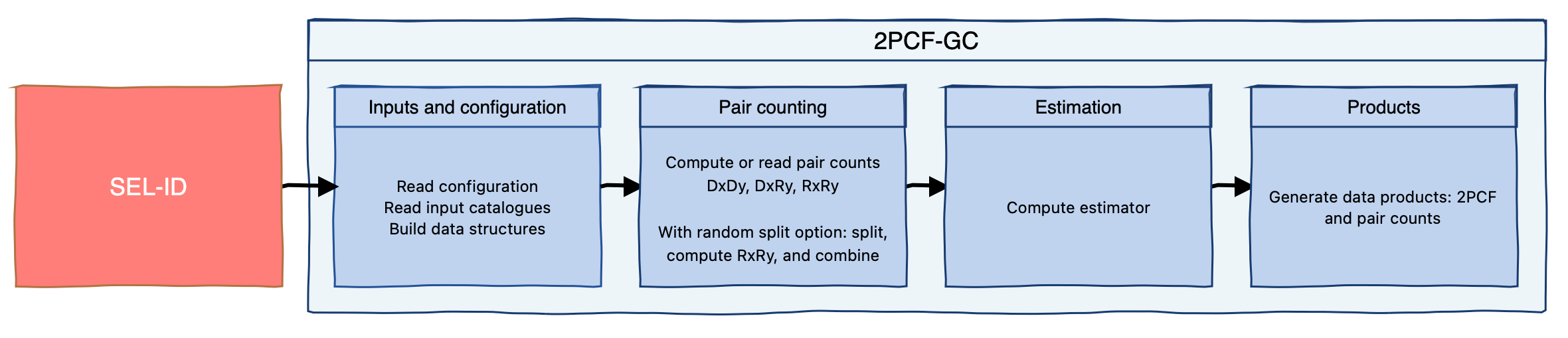}
\caption{2PCF-GC process overview. The different boxes illustrate the different steps involved in the processing.}
\label{fig:PFprocess}
\end{figure*}

The overall 2PCF-GC processing follows four main steps as illustrated in Fig. \ref{fig:PFprocess}. In the first step it reads the inputs: a configuration file, data and random catalogues, and pre-computed pair counts if this option is selected in the configuration file. In pipeline mode, those catalogues are provided by the preceding processing function in the SGS pipeline chain, the SEL-ID processing function. The latter extracts a catalogue from the Euclid Wide Survey using certain selection criteria and provides it with the associated random catalogue to 2PCF-GC. The input galaxy and random catalogues are then read and decomposed into an internal spatial representation: linked-list, kd-tree, or octree, depending on the chosen pair-counting method. After building the internal data structure, the counting algorithm roams over it to identify the pairs as a function of separation. Weighted pair counts in each separation bin are stored in arrays. 2PCF-GC performs the necessary pair counts in series depending on the requested estimator. Alternatively, they can be read from input files. Those pair counts are finally combined to obtain the 2PCF estimate. At the end of the process the 2PCF and individual pair counts products are prepared and delivered in the form of FITS files \citep{pence10}. 

\subsubsection{Inputs and outputs}

The inputs and outputs of 2PCF-GC are defined in the Euclid Common Data Model, which defines the format of the input and output data and metadata, and ensures the stability of interfaces between pipelines and the Euclid Archive System (EAS). The latter is the database containing all data products and metadata processed for the \Euclid mission \citep{williams19}. The input products of 2PCF-GC are a configuration file, a set of data and random catalogues, and possible pre-computed pair counts from a previous run. The input catalogues contain the 3-dimensional spatial information of objects as well as optional statistical weights. The latter can be used for instance to up-weight or down-weight objects in pairs to account for variations in the spatial sampling of certain objects. The input celestial coordinate system can be either the equatorial Cartesian or spherical system, where in the spherical case, the radial coordinate is either a redshift or a comoving radial distance. In the case that the redshift is provided, 2PCF-GC first converts the redshift to a comoving radial distance from a provided fiducial cosmological model. After that, Cartesian coordinates are used in the pair counting. The output products are tabulated 2PCF measurements and associated pair counts.

Apart from the input catalogues, the configuration file allows the specification of:
\begin{itemize}
    \item the correlation function estimator: auto-correlation, cross-correlation, modified auto-correlation, modified cross-correlation  
    \item the correlation function type: angle-averaged $\xi(r)$, anisotropic $\xirppi$ and $\wprp$, anisotropic $\xismu$ and $\xi_\ell$, 
    \item the pair-counting method: linked-list, kd-tree, or octree,
    \item the type of binning (linear or logarithmic) and definition of bins,
    \item the pair line-sight definition for $\xi_\ell$,
    \item the upper limit of integration along $\rpa$ for the projected correlation function,  
    \item an option to enable reusing pre-computed pair counts,
    \item the number of splits of the random catalogue when using the random split option,
\end{itemize}
through a list of parameters.
 
\section{Optimisation}

The main challenge in estimating the 2PCF from \Euclid data is to perform pair counting from massive galaxy and random catalogues, in a reasonable amount of time. \Euclid scientific accuracy requirement impose a $10\%$ accuracy relative to statistical uncertainty. This translates at estimation level into imposing a number of random catalogue objects of at least $\num{50}$ times that of the galaxy catalogue, since the estimator variance depends on the number of objects in the data and random catalogues \citep{landy93,keihanen19}. A significant effort has been put on optimising 2PCF-GC for speed and memory consumption. This was performed by making use of optimal data spatial partitioning and pair-counting method as previously discussed, but also by implementing parallelisation and a specific treatment of random-random pair counts based on a the random split technique. We describe those two aspects in the following.

\subsection{Parallelisation}

The parallelisation of pair counting can be achieved straightforwardly for the three considered algorithms. For the linked-list algorithm, this is done by splitting the loop over mesh cells in chunks and performing the computation for each chunk in parallel. In the case of the kd-tree and octree algorithms, instead of starting from the tree root, the pair counting starts in parallel from all internal tree nodes at a given depth (greater than zero). In all cases, the parallel instances computes partial counts that are then summed up in order to obtain the final counts. Those parallelisation strategies have been implemented using the shared-memory Open Multi-Processing (\texttt{OpenMP}) application programming interface. The scaling and performance of these implementations are presented in Sect. \ref{sec:valperf}.

\subsection{Treatment of random-random pair counts}

Random-random pair counts dominate the overall computation time of the 2PCF as it involves the largest number of pairs. A major gain in runtime can be obtained by using the so-called random split technique in the computation of those pairs. This technique, described in \citet{keihanen19}, relies on first splitting the random catalogue in $N_{\rm S}$ sub-catalogues, calculating all $\mathrm{DR}_i$ and $\mathrm{RR}_i$ pair counts, and finally summing up sub-catalogue pair counts to obtain the final random-random and data-random pair counts. Therefore the estimator for the 2PCF (auto-correlation case) becomes:
\begin{equation}
   \xirv = \frac{\mathrm{DD}(\br) -2 \, \mathrm{DR}'(\br) + \mathrm{RR}'(\br)}{\mathrm{RR}'(\br)}\, ,
\end{equation}
where
\begin{align}
    \mathrm{DR}'(\br) &= \frac{1}{N_{\rm S}} \, \sum_{i=1}^{i=N_{\rm S}} \mathrm{DR}_i(\br)\, , \\
    \mathrm{RR}'(\br) &= \frac{1}{N_{\rm S}} \, \sum_{i=1}^{i=N_{\rm S}} \mathrm{RR}_i(\br)\, .
\end{align}
With this strategy the maximum number of random-random pairs to be computed is smaller by a factor of $N^{-1}_{\rm S}$ compared to the case without random split, leading to a significant gain in computational time. An optimal $N_S$ is such that each random sub-catalogue is of the same size as the data catalogue. \citet{keihanen19} studied the bias and variance of the 2PCF estimator with this treatment and found that, for a random catalogue $50$ times larger than the galaxy catalogue, this technique reduces the computation time by a factor of more than ten without affecting estimator variance or bias. In the following, we adopt as baseline random catalogues that have $\num{50}$ times the number of galaxies in the data catalogue, and when using the random split technique, an optimal value of $N_S=50$. We show in the next sections that this effectively allows us to reach the accuracy requirement of \Euclid. 

\section{Validation and performance} \label{sec:valperf}

The testing and validation of 2PCF-GC have been performed at the different stages of development of the code. 2PCF-GC has successfully passed six maturity level gates, each one involving a series of validation tests including 2PCF calculations on mock data. Significant efforts have been put in making the code meet high coding standards following the quality requirement defined by SGS. We present in this section the results from the most significant validation tests, with a focus on accuracy, runtime, and memory tests.

\subsection{Benchmark catalogues}

In order to perform benchmarks of 2PCF-GC, we made use of different sets of mock data catalogues. Each one was selected to perform a specific test and has different characteristics. We review in the following those characteristics.

\begin{description}
   \item[\bf Euclid Large Mocks:] 
   the Euclid Large Mocks (ELM) suite has been designed for studying observational systematic errors on galaxy clustering. Those are realisations of H$\alpha$-emitting galaxies in a lightcone of radius 30 deg on the sky. They are based on the Pinocchio approximate method for efficiently generating halo catalogues and lightcones \citep{monaco13}. Pinocchio haloes are populated with emission-line galaxies using a halo occupation distribution model calibrated on the Flagship Galaxy Mock presented in the next section. As for the Flagship Galaxy Mock, the simulated galaxy catalogue corresponds to H$\alpha$ emitters with flux above $2\times10^{-16}~{\rm erg}\,{\rm s}^{-1}\,{\rm cm}^{-2}$ and distributed over the redshift range $0.9<z<1.8$. To perform validation tests, we only consider the galaxies at $0.9<z<1.1$ in the first mock, leading to an effective volume of $V = 1.5~h^{-3}~{\rm Gpc}^3$. Details of the ELM mocks are provided in Monaco et al. (in prep.).
   \item[\bf CoxMock suite:]
   the CoxMock suite has been designed to produce mocks for which the true 2PCF is known, and which can be produced massively in a reasonable amount of time. To do this, we used a line-point Cox process where lines of a given length are randomly placed in a cubical volume and points are randomly scattered on those lines \citep{stoyan95}. We generated $\num{10000}$ catalogue realisations of a cube of side $1.74~h^{-1}~{\rm Gpc}$ containing $10^6$ objects. For each realisation, $\num{100000}$ random lines of length $800\, h^{-1} \, {\rm Mpc}$ were drawn. The expected 2PCF of points in these mocks resembles a cosmological $\xi(r)$ as it is described by a damped power-law with an index of $-2$ (see Sect. \ref{subsec:accuracy}).
\end{description}

\subsection{Runtimes and scaling with number of objects}

We tested the runtime of 2PCF-GC with the different pair-counting algorithms and as a function of the number of objects in the catalogues. For this, we extracted four sub-catalogues with $N_{\rm o}=[10^4, 10^5, 10^6, 5\times10^6]$ randomly chosen objects from the ELM mock. We then built the associated random catalogues with $\num{50}$ times more objects than in the data. The redshift distribution of random objects was drawn from the estimated mean distribution over \num{1000} mocks. Since the volume of the mock is the same, the sub-catalogues probe different number densities of objects. In this test, we considered the estimation of the multipole correlation functions in $\num{40}$ linear bins in $r$ spanning the interval $[0, 200]~\si{\hMpc}$ and $\num{200}$ in $\mu$. The runtimes are obtained using a computer cluster node equipped with $\num{32}$ physical central processing units (CPU) Intel(R) Xeon(R) Silver 4216 at 2.10 GHz, and by performing $\num{5}$ runs with $\num{4}$, $\num{8}$, $\num{16}$, or $\num{32}$ parallel threads, except for the longest runs with $N_{\rm o}>10^6$ where only one run with $\num{32}$ threads was performed.

\begin{figure}[htbp!]
\centering
\includegraphics[angle=0,width=\hsize] {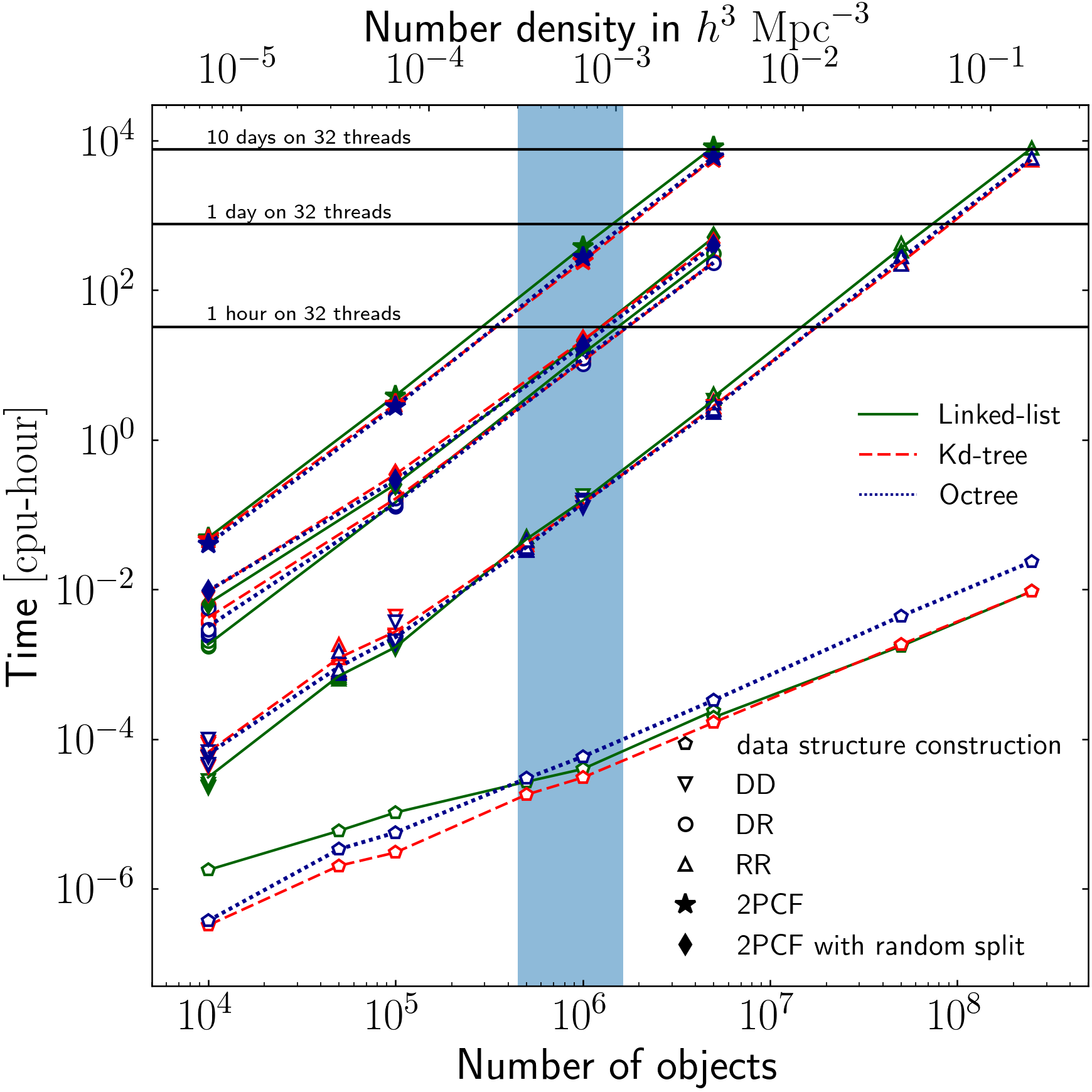}
\caption{Runtimes for the calculation of the multipole correlation of galaxies obtained from the ELM mock. The times are expressed in CPU-hour and as a function of the data or random catalogue size. The various symbols represent the time spent on the data structure construction, DD calculation, DR calculation, RR calculation, and the overall 2PCF runtime with and without random split option. The DR calculation times are provided as a function of the number of objects in the data catalogue and assuming a fifty times larger random catalogue. The different curves represent the runtimes obtained with the linked-list (solid), kd-tree (dashed), and octree (dotted) algorithms. The blue vertical band shows the range of expected number densities in the spectroscopic sample at redshifts within $0.9<z<1.8$. The abscissa refers to the number of object in the data catalogue except for RR calculation where it refers to the that in the random catalogue.}
\label{fig:runtimes}
\end{figure}

Figure \ref{fig:runtimes} presents the times in CPU-hour that we obtained when estimating the multipole correlation functions with the different algorithms, as a function of the number of objects. The total runtimes, denoted by 2PCF in the figure, are divided in different parts corresponding to the time spent on data structure construction, DD and RR pair counting. While the data structure construction time scales linearly with the number of objects, the DD, DR, and RR pair-counting times scale quadratically for a constant volume, as expected. Overall, the data structure construction has a subdominant contribution of one up to seven orders of magnitude smaller than pair counting. The DD, DR, and RR pair-counting times are similar for the tree-based algorithms, while the linked-list algorithm shows a steeper slope with the number of objects, that is, smaller times for low $N_{\rm o}$ and slightly longer times above $10^6$ objects. The full calculation of the 2PCF is dominated by the RR counts and the overall runtimes reflect this, with the tree-based algorithms and particularly the octree performing best. Furthermore, the random split option allows us to drastically reduce the runtimes by up to a factor of $\num{10}$ for a data catalogue of $5\times10^6$ objects, as shown in Fig. \ref{fig:runtimes}. In that case, a data catalogue is processed in a little less than one day on $\num{32}$ threads, while without this lasts $\num{8}$ days. The scaling with the number of objects is slightly different for the three pair-counting algorithms. Since the individual random split sub-catalogues have approximately the same size as the data catalogue, the times are dominated by smaller-size catalogues for which the linked-list algorithm proves to be slightly faster. This is true up to approximately $N_{\rm o}=5\times10^5$, and at larger $N_{\rm o}$ the runtimes are similar for the three algorithms, with only the octree being marginally faster for $5\times10^6$ objects. For the full 2PCF calculation, the linked-list algorithm runtimes scale very accurately as $N^2_{\rm o}$, while for tree-based algorithms as $N^{1.96}_{\rm o}$ in the regime of $N_{\rm o}>10^5$.

The volume of the ELM mock that we used for runtime tests corresponds approximately to the volume probed spectroscopically by the Data Release 1 of \Euclid in the redshift interval $0.9<z<1.1$. We show with the vertical blue band in Fig. \ref{fig:runtimes} the expected number density of galaxies over $0.9<z<1.8$, which gives an idea of the expected runtime to measure the multipole correlation functions in the first redshift interval: about $50$min on $\num{32}$ threads. It is worth noting that the runtimes also scale in an independent manner with volume and maximum scale. Up to approximately $\num{5}$ millions of $\mathrm{H}\alpha$-emitting galaxies $0.9 < z < 1.1$ are expected in the spectroscopic catalogue by the end of the Euclid Wide Survey, and according to our tests, the calculation would last a bit less than a day on $\num{32}$ threads using the random split option. This represents a reasonable time and meets \Euclid mission requirement, given that more than $\num{32}$ threads will be usable in the SDC for the actual \Euclid data processing. In particular, some SDC can provide up to $\num{128}$ usable parallel threads.

\subsection{Accuracy tests} \label{subsec:accuracy}

Within the validation of 2PCF-GC, we conducted a series of tests on the accuracy of the estimated 2PCF. We tested both the absolute accuracy and relative accuracy to estimates obtained from an external software. Those tests are presented in the following.

\subsubsection{Absolute accuracy}

We made a comprehensive analysis that uses a series of mocks for which the underlying 2PCF is perfectly known, namely the CoxMock suite of mocks. We generated $\num{10000}$ realisations allowing us to reach extremely precise summary statistics on the 2PCF. The reference bin-averaged CoxMock (isotropic) correlation function is given by \citep{stoyan1987stochastic}
\begin{equation}
    \xi_{\rm ref}(r_{\rm low},r_{\rm up}) = \frac{3}{4 \pi L^2 n_{\rm L}} \frac{(r_{\rm up}-r_{\rm low})(2L - r_{\rm up} - r_{\rm low})}{r^3_{\rm up}-r^3_{\rm low}} - \frac{1}{N_{\rm L}}\, ,
\end{equation}
where $r_{\rm low}$ and $r_{\rm up}$ are the lower and upper limits of the bin in $r$, $L$ is the line length, $n_{\rm L}$ is the line number density, and $N_{\rm L}$ is the number of lines. The last term is a correction term in the Cox correlation function that accounts for using a finite number of lines.

\begin{figure}[htbp!]
\centering
\includegraphics[angle=0,width=\hsize] {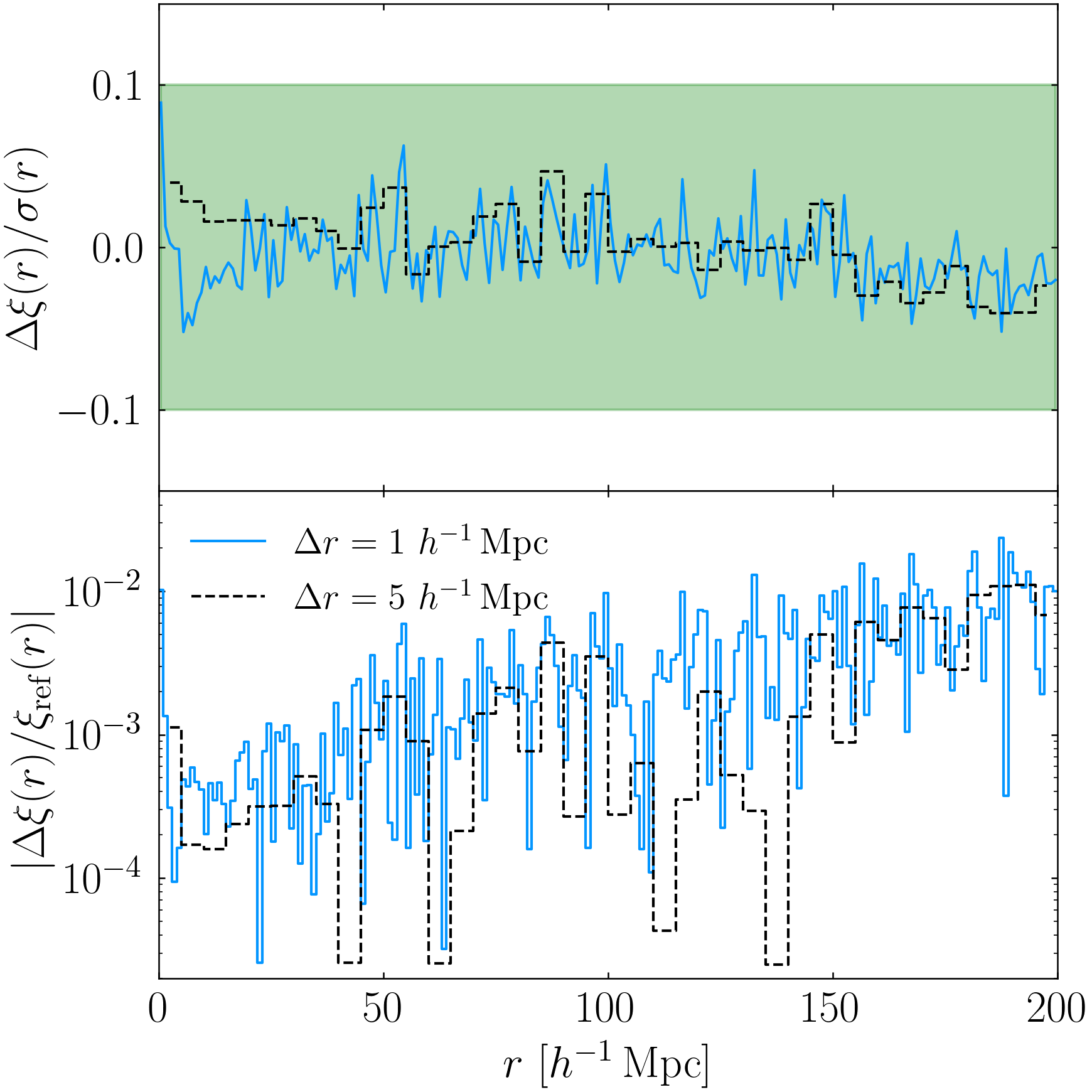}
\caption{Accuracy of the estimated real-space correlation function from the CoxMocks. While the top panel presents the mean bias relative to the standard deviation $\Delta \xi/\sigma$, the bottom panel exhibits the absolute value of the mean relative bias $|\Delta \xi/\xi_{\rm ref}|$. These estimates are obtained by averaging over \num{10000} mock realisations.}
\label{fig:coxxi}
\end{figure}

We measured the angle-averaged and multipole correlation functions in the mock realisations using $200$ and $40$ linear bins in $r$ spanning the range $[0,200]$ $\, h^{-1} \, {\rm Mpc}$ and including the random split option. In those mocks there is no anisotropic clustering and we recover vanishing quadrupole and hexadecapole multipole moments with residual random variation around zero, as expected. To quantify the accuracy on the estimation, we calculate the mean bias relative to the standard deviation $\Delta \xi/\sigma$ and mean relative bias $\Delta \xi/\xi_{\rm ref}$, respectively defined as 
\begin{align}
    \frac{\Delta \xi(r)}{\sigma(r)} &= \frac{\langle\xi(r)\rangle - \xi_{\rm ref}(r)}{\sigma(r)}\, , \\
    \frac{\Delta \xi(r)}{\xi_{\rm ref}(r)} &= \frac{\langle\xi(r)\rangle - \xi_{\rm ref}(r)}{\xi_{\rm ref}(r)}\, ,
\end{align}
where $\langle\xi(r)\rangle$ refers to the mean $\xi(r)$ over the realisations, and $\sigma$ corresponds to the reference statistical uncertainty in \Euclid. The latter, which varies between $2\times10^{-3}$ and $9\times10^{-5}$ at $r$ scales below $200\, h^{-1} \, {\rm Mpc}$, was derived in \citet{Blanchard-EP7} under the assumption of Gaussian covariances. The mean over the $\num{10000}$ realisations suppresses field stochasticity and sample variance. Both quantities are shown in Fig. \ref{fig:coxxi}. We find that the mean bias in the estimated correlation is extremely low, always below 10\% of the statistical uncertainties for all considered scales. There are residual sample variance fluctuations, but overall we do not find any trend of bias. On closer inspection, we find that, except on scales below $10\, h^{-1} \, {\rm Mpc}$, the mean bias is always below about 5\%. Similarly, the mean relative bias is always below $1\%$ and can reach $0.1$ per cent at the smallest scales.   

\subsubsection{Relative accuracy}

\begin{figure}[htbp!]
\centering
\includegraphics[angle=0,width=\hsize] {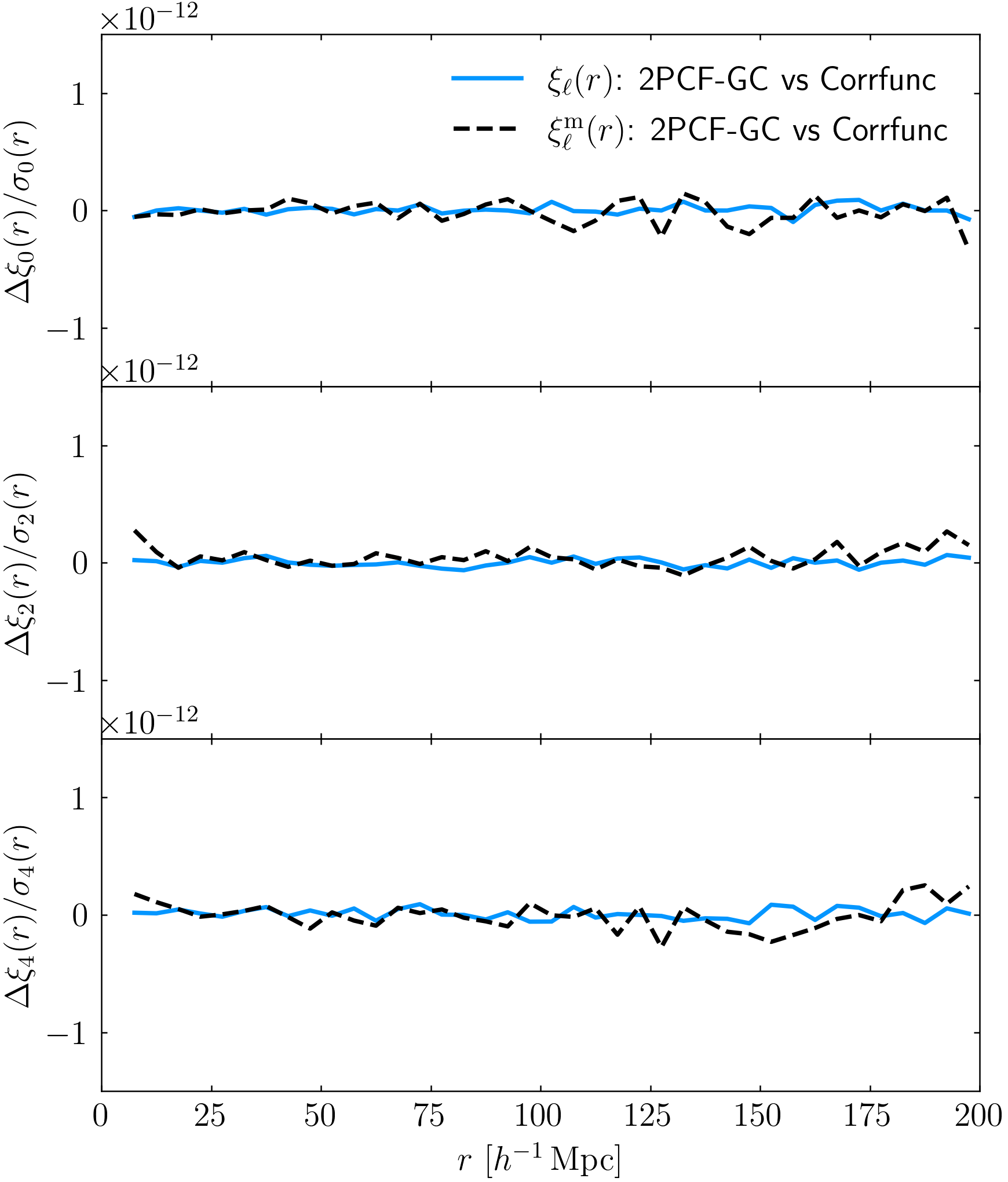}
\caption{Difference in the monopole (top panel), quadupole (central panel), and hexadecapole (bottom panel) moments of the standard ($\xi_\ell$) and modified ($\xi^{\rm m}_\ell$) correlation function between the 2PCF-GC and \texttt{Corrfunc} estimates, relative to the standard deviation. These measurements are obtained from a single mock realisation.}
\label{fig:pycorr}
\end{figure}

We further compared the standard and modified correlation function multipole moments measured by 2PCF-GC with those obtained with the publicly available \texttt{Corrfunc} code \citep{sinha20}. For this test we used the Flagship Galaxy Mock v1 galaxies at $0.9<z<1.1$ (see Sect. \ref{sec:expect}) and considered $40$ linear bins in $r$ spanning the range $[0,200]$ $\, h^{-1} \, {\rm Mpc}$. The difference in the monopole, quadrupole, and hexadecapole relative to the expected statistical error expected in \Euclid, are presented in Fig. \ref{fig:pycorr}. For each multipole moment, the relative accuracy between the two estimates is of the order of the machine precision, both for the standard and modified 2PCF. In the modified 2PCF estimation, the reconstructed data and auxiliary random catalogues were obtained using the recSym BAO reconstruction algorithm, as detailed in Sarpa et al (in prep.). 

\begin{figure}[htbp!]
\centering
\includegraphics[angle=0,width=\hsize] {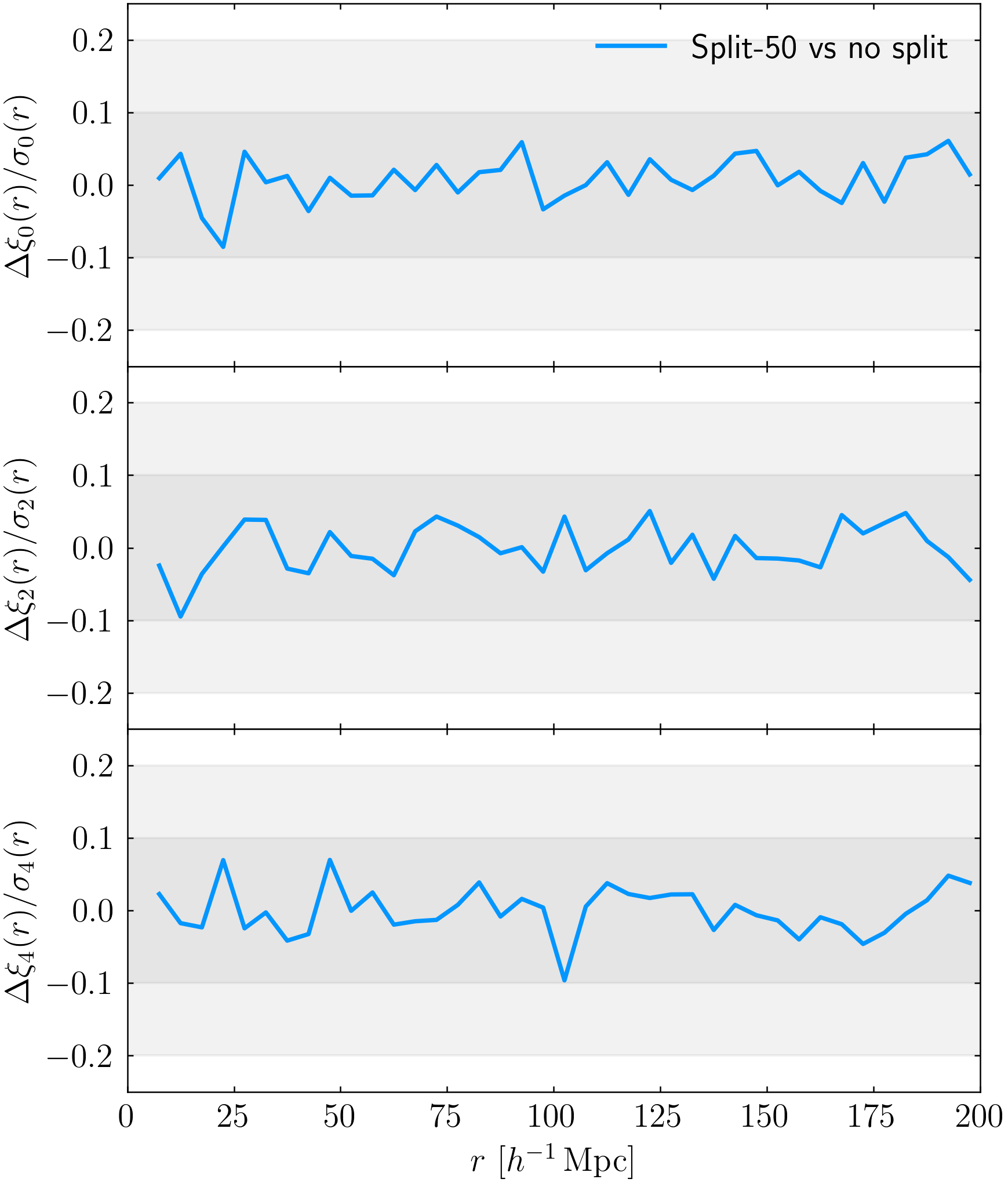}
\caption{Impact of the random split option on the accuracy of the measured 2PCF. The curve shows the difference in  in the monopole (top panel), quadupole (central panel), and hexadecapole (bottom panel) moments of the correlation function, while using or not the random split option with $N_{\rm S}=50$. The shaded areas represent the regions encompassing respectively 10\% and 20\% of the expected statistical uncertainty. These measurements are obtained from a single mock realisation.}
\label{fig:split}
\end{figure}

Finally, we tested the impact of using the random split option on the accuracy of the estimated 2PCF multipole moments. For this, we compared the measurements obtained in the same mock catalogue with the same setup as before while using or not this option. The difference in the monopole, quadrupole, and hexadecapole relative to the expected statistical error is shown in Fig. \ref{fig:split}.
The results show that the random split option with $N_{\rm S}=50$ introduces random errors below 10\% of the statistical uncertainties for all considered scales and no systematic error. It is worth emphasising that those uncertainties can be reduced by reducing the number of random splits, at the expense of longer runtimes.

Overall, these tests validates 2PCF-GC accuracy and the choice of using $\num{50}$ times more random points than objects in the data catalogue and the usage of the random split option. We would like to emphasise that, even if not presented here, we also performed other accuracy and consistency tests on the cross-correlation function, and in all cases, we found similar, very good accuracies. This includes for instance consistency tests of the cross-correlation function where the two data catalogues are taken as two random sub-samples of a data catalogue. 2PCF-GC has been used to estimate cross-correlation functions in Risso et al. (in prep.).

\subsection{Memory usage}

The random-access memory (RAM) usage of 2PCF-GC is dominated by the storage of the input data and random object positions and weights. Furthermore, the memory required to process the data can increase depending on the amount of information comprised by the linked-list or tree structures. Although object properties are not directly copied into the spatial partitioning structures, additional metadata associated with the latter must be stored in memory. The linked-list structure, which comprises an ordered array of integers that convey information about the objects present in each cell, formally requires the least memory. The tree structures generally require the storage of more information. Each tree node contains several pieces of information including the coordinates of the bounding volume associated with the node, the indices of the objects contained in the node, and pointers to child nodes. This increases the amount of information to be stored in memory. However, the use of hash tables allows the pointer information to be removed, and in the case of the octree, the Morton code associated with each node allows the bounding volume coordinates to be computed directly, thus saving further memory.

\begin{figure}[htbp!]
\centering
\includegraphics[angle=0,width=\hsize] {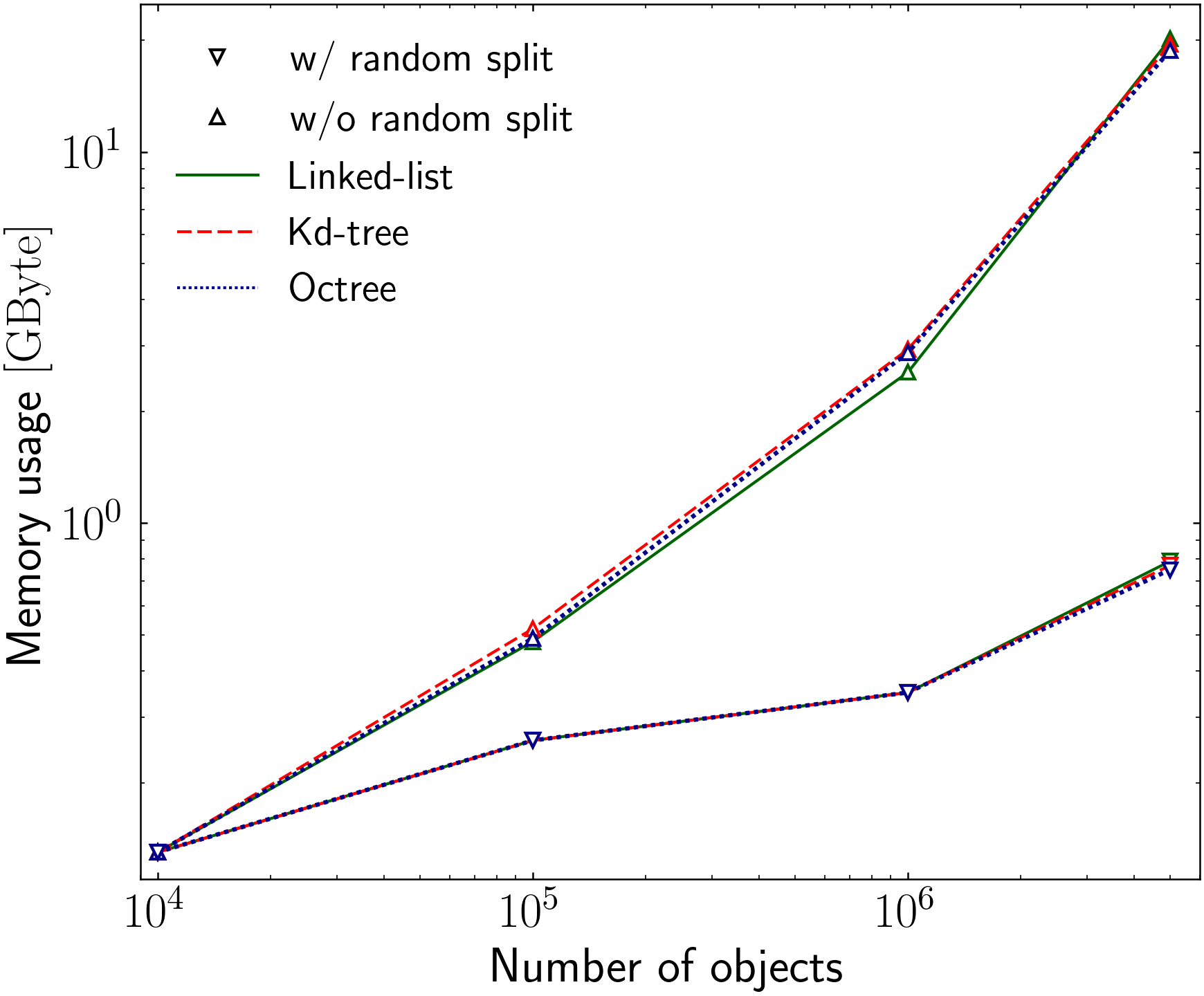}
\caption{Memory usage as function of the number of objects in the data catalogue, for the considered data partitionings and while using or not the random split option. The different lines shows the results for the different pair-counting methods, while the different symbols indicate whether or not the random split option is used.}
\label{fig:ram}
\end{figure}

In Fig. \ref{fig:ram} we present estimates of the memory usage of 2PCF-GC from the ELM mock, both accounting for the RAM resident set size and swap space, for the three methods and as a function of the number of objects in the data catalogue. The memory usage increases with the number of objects and the storage of the input catalogues dominates the overall budget.  The linked-list method tends to use less memory, but thanks to the implementation of hash tables for both kd-tree and octree, the memory usage is only marginally larger for the latter methods. Overall, for a data catalogue of five million objects and 50 times more in the random catalogue, about $19~\mathrm{GB}$ of memory is required. In the case where the random split option is used, only a fraction of the random catalogue needs to be stored in memory at a time and thus the necessary memory for storing the random object information is significantly lower. In that case only about $0.9~\mathrm{GB}$ of memory is required for a data catalogue containing five million objects.

\section{Expectations for the Euclid Wide Survey} \label{sec:expect}

We present in this section some forecasts on 2PCF estimation and showcase some interesting features of 2PCF-GC using a realistic full-sky mock catalogue of the Euclid Wide Survey. We made use of the Flagship Galaxy Mock (FGM) v2.1 \citep{carretero17,tallada20}, which has been constructed from the Flagship 2 N-body simulation by populating dark matter haloes with emission-line galaxies as described in \citet{EuclidSkyFlagship}. A sub-sample of the galaxies with an $\mathrm{H}\alpha$ flux greater than $2\times10^{-16}~{\rm erg}\,{\rm s}^{-1}\,{\rm cm}^{-2}$ and a redshift in the range $0.9<z<1.8$ has been extracted, which represents the targeted galaxies for the spectroscopic sample of the Euclid Wide Survey. The mock consist of a light-cone covering an octant of the sky, and for the purpose of performing realistic forecasts, we replicated the octant in order to have full-sky catalogue. We note that this procedure does not introduce spurious clustering features on the scales smaller than $200~h^{-1}~\mathrm{Mpc}$, which are of interest for our analysis. This mock represents current most realistic expectation for the intrinsic clustering of H$\alpha$-emitting galaxies in the Euclid Wide Survey available within the \Euclid collaboration.

\subsection{Galaxy 2-point correlation function across survey timeline}

\begin{figure}[htbp!]
\centering
\includegraphics[angle=0,width=\hsize] {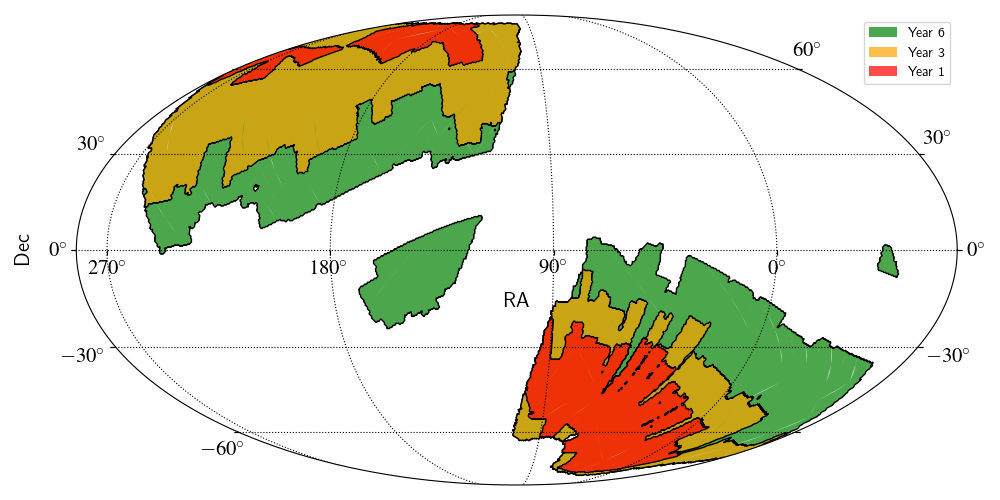}
\caption{Sky coverage of Euclid Wide Field after one, three, and six years of observations, as defined in the Euclid Reference Survey \citep{Scaramella-EP1}.}
\label{fig:footprint}
\end{figure}

To illustrate the performance of 2PCF-GC and provide forecasts on the expected measured 2PCF at different stages of \Euclid observations, we estimated the correlation function multipole moments in the FGM after 1, 3, and 6 years of observations, as defined in the Euclid Reference Survey\footnote{In this analysis we use the Euclid Reference Survey rsd2024A.} \citep{Scaramella-EP1}. The angular extent of year 1, year 3, and year 6 observations as defined in the Euclid Reference Survey is presented in Fig. \ref{fig:footprint}. We defined four redshift intervals: $0.9<z<1.1$, $1.1<z<1.3$, $1.3<z<1.5$, $1.5<z<1.8$, as used for \Euclid cosmological forecasts \citep[e.g.,][]{Blanchard-EP7}.

\begin{figure}[htbp!]
\centering
\includegraphics[angle=0,width=\hsize] {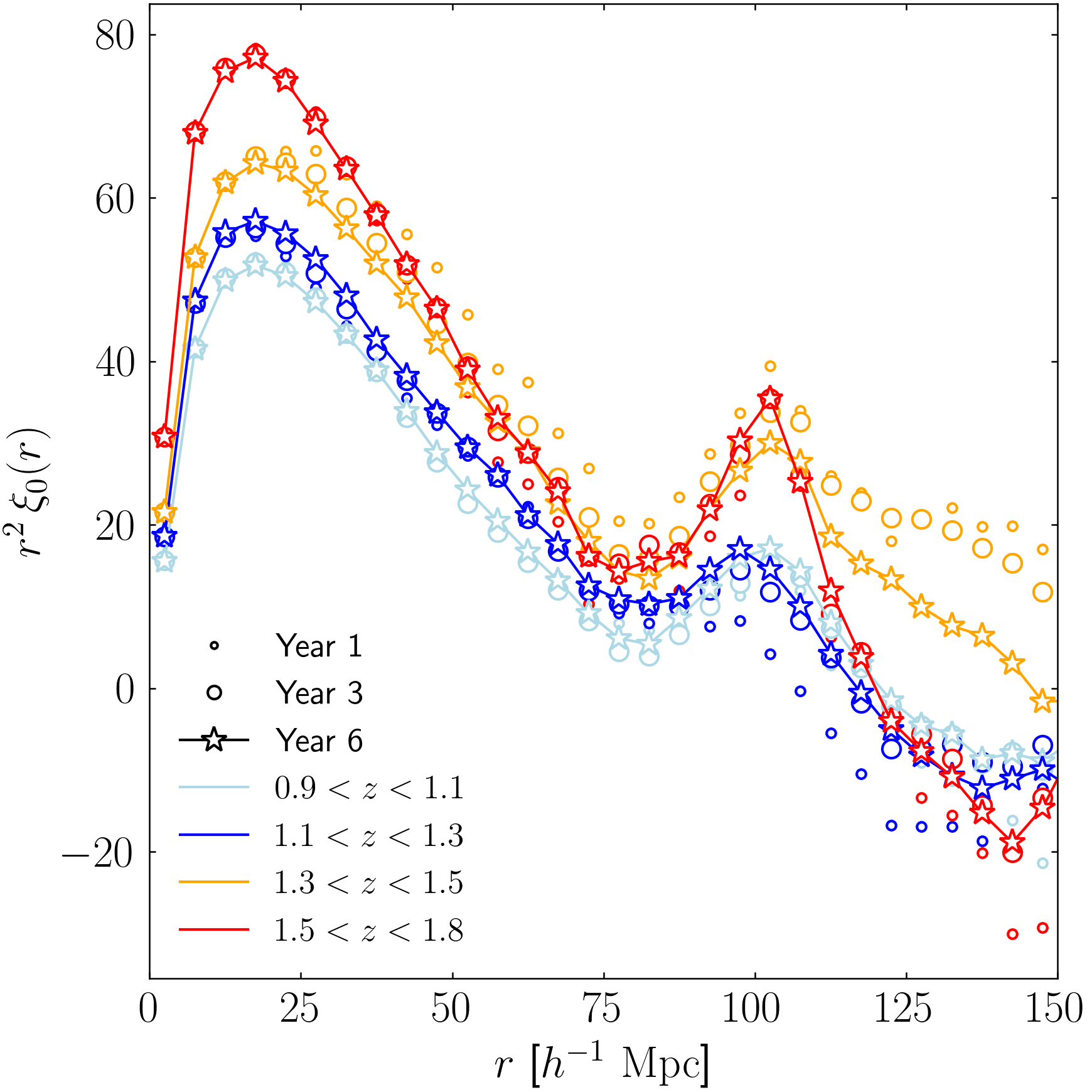}
\caption{Monopole correlation function estimated from the FGM mock for galaxies with $\mathrm{H}\alpha$ flux above $2\times10^{-16}~\mathrm{erg}~ \mathrm{s}^{-1}~\mathrm{cm}^{-2}$ at different epochs of observations. The different colours show the monopole in the redshift intervals: $0.9<z<1.1$, $1.1<z<1.3$, $1.3<z<1.5$,$1.5<z<1.8$.}
\label{fig:xi0}
\end{figure}

\begin{figure}[htbp!]
\centering
\includegraphics[angle=0,width=\hsize] {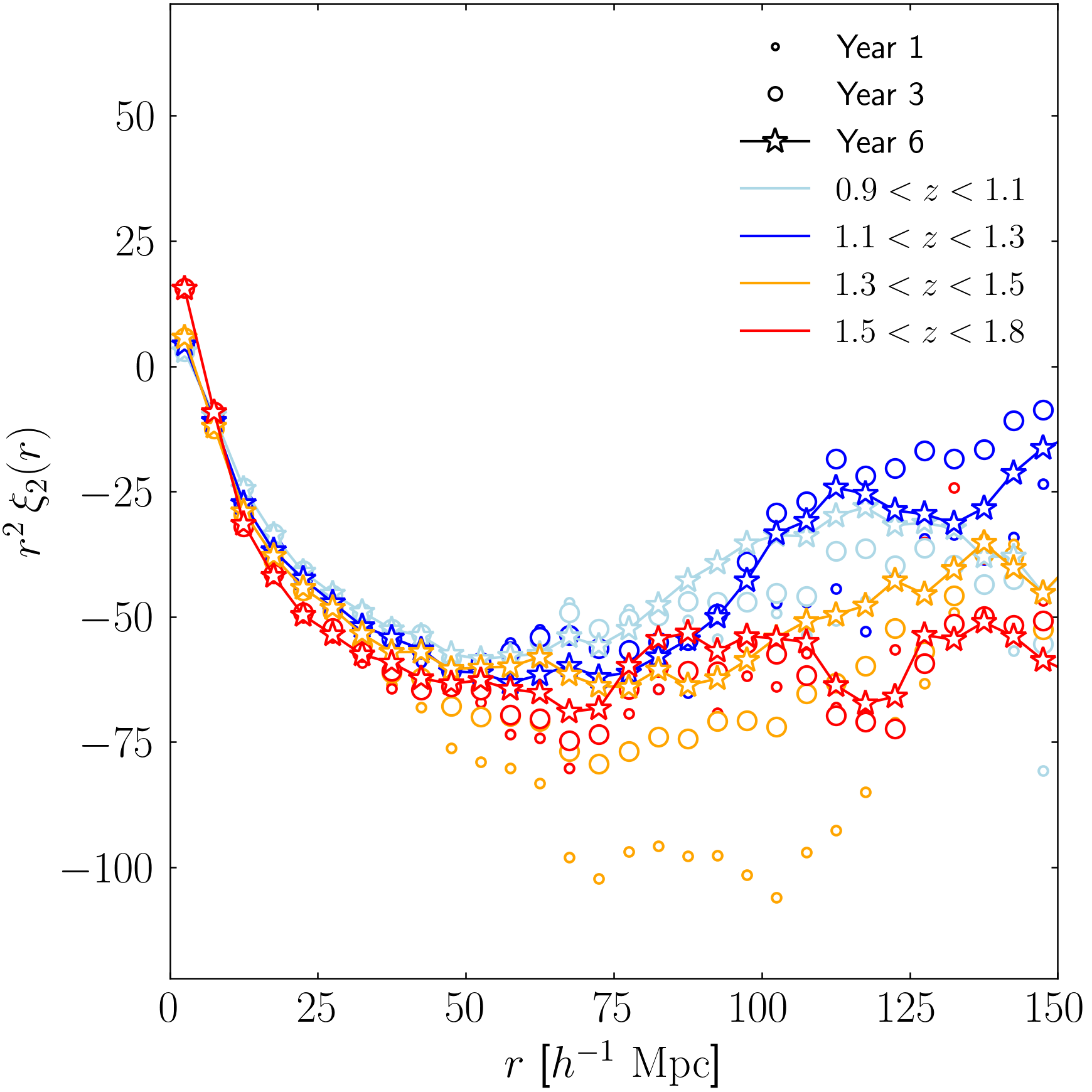}
\caption{Same as Fig. \ref{fig:xi0} but for the quadrupole correlation function.}
\label{fig:xi2}
\end{figure}

\begin{figure}[htbp!]
\centering
\includegraphics[angle=0,width=\hsize] {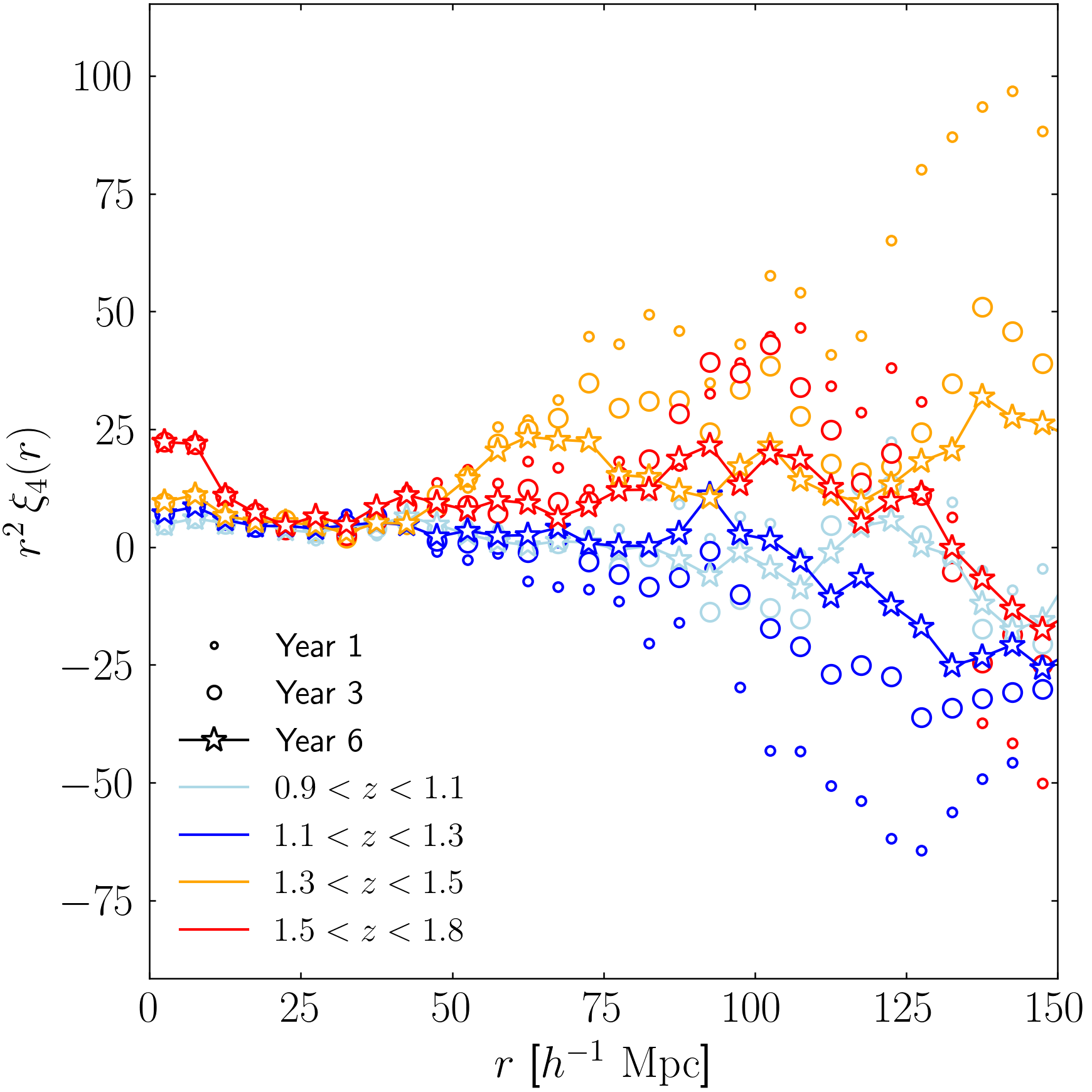}
\caption{Same as Fig. \ref{fig:xi0} but for the hexadecapole correlation function.}
\label{fig:xi4}
\end{figure}

The correlation function monopole, quadrupole, and hexadecapole are presented in Fig. \ref{fig:xi0}, Fig. \ref{fig:xi2}, and Fig. \ref{fig:xi4} respectively. Even if these measurements are extracted from a specific mock realisation of the observable universe, with its own sample variance, it gives a sense of the improvement on the estimation of the correlation function and high precision recovered after 6 years of observation. In particular, one can see from the monopole that the BAO peak in the lowest redshift interval is not very pronounced with the first year coverage, while at the end of the survey, the signal is much more significant. On scales beyond $100~h^{-1}~\mathrm{Mpc}$, we can see significant variations of the amplitude between year 1 and year 6, which can be attributed to sample variance, the fact that the overdensity distribution associated with galaxies differs significantly. Similar trends are seen in the quadrupole and hexadecapole. Particularly for the latter, which is the most uncertain, one can see that the end-of-survey amplitude tends to vanish on large scales, as expected.

\subsection{Impact of pair line-of-sight definition}

\begin{figure}[htbp!]
\centering
\includegraphics[angle=0,width=\hsize] {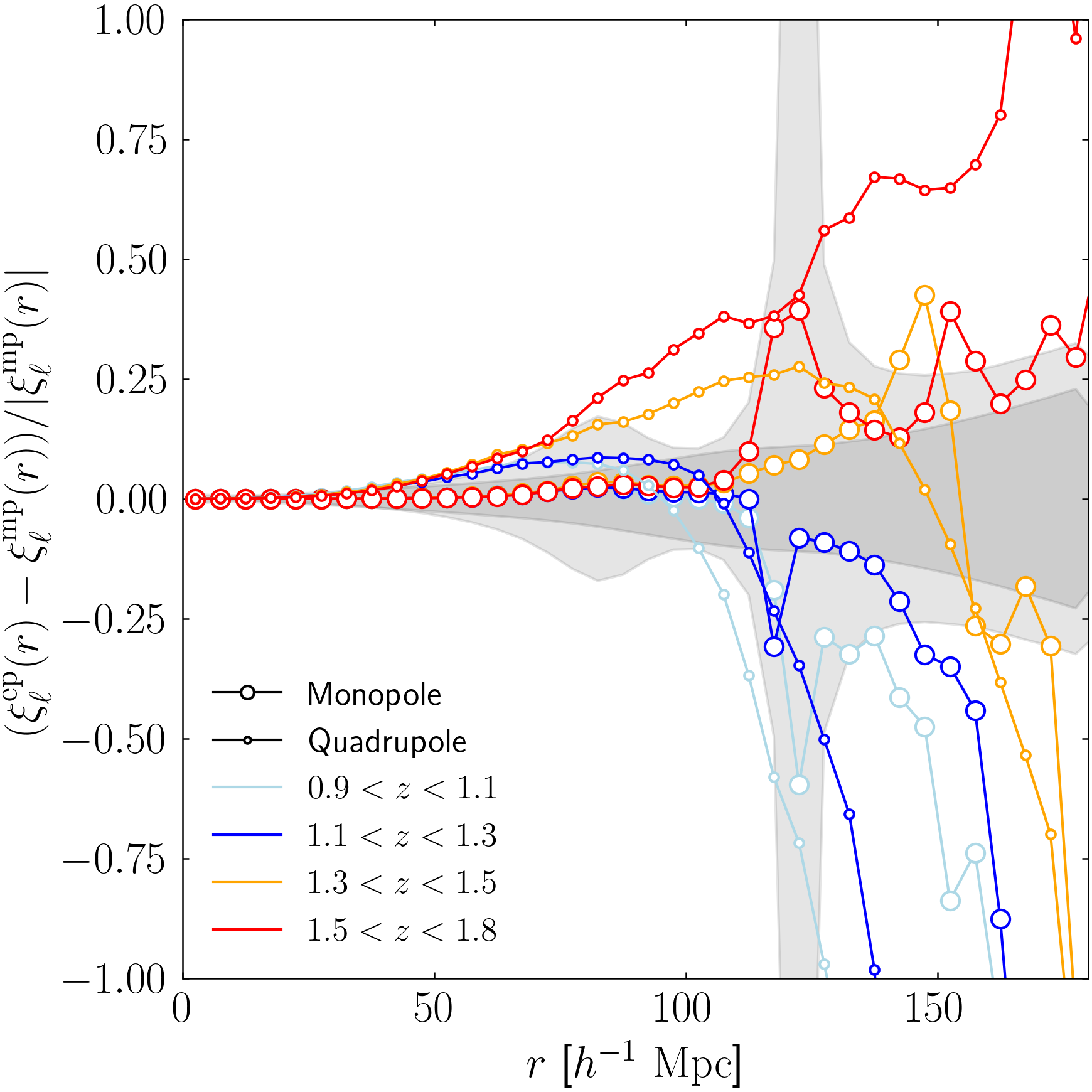}
\caption{Impact of the choice of pair line-of-sight definition on the estimated monopole and quadrupole correlation functions in FGM mock. The curves with the different different symbols show the relative difference between using mid-point and end-point definitions in the estimator for the monopole and quadrupole correlation functions. The different colours show this quantity for the redshift intervals: $0.9<z<1.1$, $1.1<z<1.3$,$1.3<z<1.5$,$1.5<z<1.8$. The light (dark) band shows the expected $1\sigma$ statistical error on the monopole (quadrupole) correlation function in the completed Euclid Wide Survey.}
\label{fig:xilos}
\end{figure}

In order to showcase the capabilities of 2PCF-GC, we estimated the multipole correlation function in the FGM using the mid-point and end-point line-of-sight definition. Even if the estimator assumes the local plane-parallel approximation, the choice of the line-of-sight definition has different sensitivity to wide-angle effects. From its maximally-symmetric properties, the mid-point definition minimises the latter effects, while the end-point definition is more affected \citep[e.g.,][]{reimberg16,beutler19}. This can be seen in Fig. \ref{fig:xilos}, which shows the relative difference between using the mid-point and end-point definitions in the estimator, for the monopole and quadrupole correlation functions in the year 6 data set. In particular, it is instructive to see a relative difference of less that $5\%$ ($25\%$) appears below (above) $100\, h^{-1} \, {\rm Mpc}$ with no strong redshift dependence in the monopole. This difference remains with the  $1\sigma$ statistical error expected in the Euclid Wide Survey shown with the dark green bands in the figure.
In the quadrupole instead, the effect can reach up to $25\%$ at $100\, h^{-1} \, {\rm Mpc}$ in the last two redshift intervals, at $1.3<z<1.5$ and $1.5<z<1.8$, while it is of less than $10\%$ at the lowest redshifts. In the two highest redshift intervals, the difference goes beyond the expected statistical error in the quadrupole. Overall, these results gives a lower limit on the typical wide-angle effects expected on large scales in the final Euclid Wide Survey sample, and arising from the use of the plane-parallel approximation in the estimator.

\subsection{Integral constraint}

The normalisation of the pair counts to the observed number of galaxies in the LS estimator imposes a constraint on the integral of the observed overdensity in the survey. However, because of the finite size of the sample, the integral over the observed overdensity field does not necessarily vanish and this leads to biasing negatively the correlation function on large scales \citep{peebles80}. This effect, commonly referred to as the integral constraint, is particular important for small surveys. In the Euclid Wide Survey, which will be among the largest spectroscopic survey of galaxies, the expected effect is very small. Nonetheless, given the unprecedented statistical precision in the correlation function measurements expected in \Euclid, it is important to assess the level and scales at which the integral constraint can impact the measurements.     

The integral constraint introduces a bias on the observed correlation function such that the latter relates to the true correlation function as
\begin{equation} \label{eq:xiic}
    \xi_{\rm obs}(\Vec{r}) \simeq \xi(\Vec{r}) - \langle \epsilon^{2} \rangle,
\end{equation}
where $\epsilon = \int_{\rm V} \diff^3 x \, \delta (\Vec{x}) \, W(\Vec{x})~\bigg/ \int_{\rm V}  \diff^3 x \, W(\Vec{x})$,  $\delta (\Vec{x})$ is the overdensity field, $W(\Vec{x})$ is the survey window function, and $\int_{\rm V} \diff^3 x$ denotes an integral over the survey volume. The leading integral-constraint term in the correlation function is \citep[e.g.,][]{demattia19}
\begin{align} \label{eq:ic}
    \langle \epsilon^{2} \rangle &= \frac{\int_{\rm V} \diff^3 \Delta \, \xi(\Vec{\Delta}) \mathcal{W}(\Vec{\Delta})}{\int_{\rm V} \diff^3 \Delta \, \mathcal{W}(\Vec{\Delta})}
    = \frac{\int_0^{\infty} \diff\Delta \, \Delta^2 \sum_{\ell=0}^{\infty} \frac{\xi_\ell(\Delta)}{2\ell + 1} \mathcal{W}_\ell(\Delta)}{\int_0^{\infty} \diff\Delta \, \Delta^2 \, \mathcal{W}_0(\Delta)},
\end{align}
where the survey window function correlation function and associated multipole moments are
\begin{align}
    \mathcal{W}(\Vec{\Delta}) &= \int_{\rm V} \diff^3 x \, W(\Vec{x}) W(\Vec{x}+\Vec{\Delta}), \\
    \mathcal{W}_\ell(\Delta) &= \frac{2\ell + 1}{2} \int_{-1}^{1} \diff \mu_\Delta \mathcal{W}(\Vec{\Delta}) \, L_\ell(\mu_\Delta),
\end{align}
respectively. In the previous equations, $\Vec{\Delta}$ denotes a separation vector. By expanding Eq. \eqref{eq:xiic} in multipole moments, one can see that the constant integral constraint term will only survive in the monopole correlation function. Moreover, by realising that the survey window function correlation function is directly related to random-random counts by $\mathrm{RR}(\Vec{\Delta}) \propto \Delta^2 \mathcal{W}(\Vec{\Delta})$ \citep{wilson17}, Eq. \eqref{eq:ic} can be rewritten as
\begin{align}
  \langle \epsilon^{2} \rangle &= \frac{\int_0^{\infty} \diff\Delta \, \sum_{\ell=0}^{\infty} \, (2\ell + 1)^{-1} \mathrm{RR}_\ell(\Delta) \xi_\ell(\Delta) }{\int \diff\Delta \, \mathrm{RR}_0(\Delta)} \\
  &= \frac{\sum_{i=1}^{N_b} \, \sum_{\ell=0}^{\infty} \,  (2\ell + 1)^{-1}  \, \mathrm{RR}_\ell(\Delta_i) \, \xi_\ell(\Delta_i)}{ \sum_{i=1}^{N_b} \, \mathrm{RR}_0(\Delta_i)}, \label{eq:icfinal}
\end{align}
where in Eq. \eqref{eq:icfinal} the continuous integral is approximated by a discrete sum over bins\footnote{Formally, here the bins in $\Delta$ need to be equally spaced.} $\Delta_i$, and $N_b$ is the number of bins. In the previous equation, $\mathrm{RR}_\ell$ is defined as
\begin{equation}
\mathrm{RR}_\ell(r)=\frac{2\ell+1}{2} \int_{-1}^{1}  \diff\mu \, \mathrm{RR}(s,\mu) \, L_\ell(\mu)\, .   
\end{equation}
Therefore, by using Eq. \eqref{eq:icfinal} we can directly estimate the impact of the integral constraint from the random-random pair counts and a prior knowledge on the true underlying multipole correlation function. 

We compute the integral constraint correction expected in \Euclid from  Eq. \eqref{eq:icfinal} by using the $\mathrm{RR}$ counts estimated by 2PCF-GC and performing the discrete sum up to $\Delta=1000~ h^{-1}~\mathrm{Mpc}$. The correlation function multipoles in the integral are taken from the measured multipole (symbols) or by using an analytical model for the redshift-space multipole correlation function. The latter was obtained by fitting the FGM measurements with most accurate full-shape models for the correlation function multipole moments, as described in \citet{karcher24}. The main advantages of using a model $\xi_\ell$ is to have a prediction up to the largest scales and that is not itself affected by the integral constraint. Overall, we found values of $\langle \epsilon^{2} \rangle$ varying within (3--8)$\times 10^{-5}$, depending on the epoch of observations and the redshift interval.

\begin{figure}[htbp!]
\centering
\includegraphics[angle=0,width=\hsize] {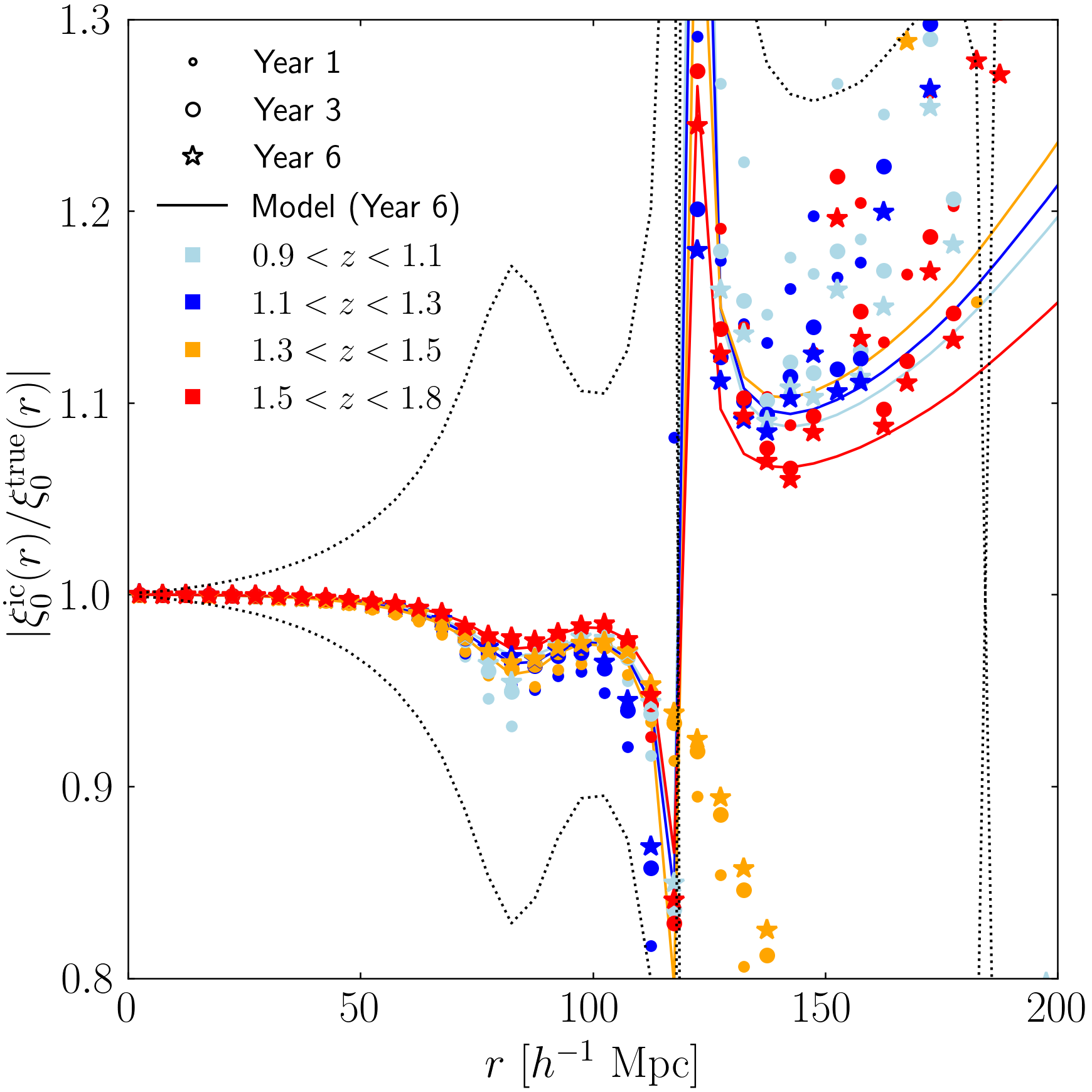}
\caption{Impact of the integral constraint on the monopole correlation function in the Euclid Wide Survey. The curves with the different different symbols show the absolute value of the ratio between the monopole correlation functions affected by integral constraint and true underlying one, after 1, 3, and 6 years of observation respectively. The solid curves show the same quantity for year 6 observations, but when the integral constraint effect is derived from a model correlation function. The different colours show this quantity for the redshift intervals: $0.9<z<1.1$, $1.1<z<1.3$,$1.3<z<1.5$,$1.5<z<1.8$. The dotted curves delineate the expected $1\sigma$ statistical uncertainty on the monopole correlation function in the completed Euclid Wide Survey.}
\label{fig:ic}
\end{figure}

We present in Fig. \ref{fig:ic} the absolute value of the ratio between the observed monopole correlation function from the FGM and the integral constraint-corrected (or true) one, for the different redshift intervals and observing periods considered. We can see that the impact of the integral constraint depends on the size of the surveyed volume and the considered redshift interval. At year 1, the effect reaches about 10 percent at the BAO scale, at $r=100~h^{-1}~\mathrm{Mpc}$, and up to 20--30$\%$ at $r=200~h^{-1}~\mathrm{Mpc}$. The effect is the most prominent for the lowest redshift interval. At year 3 and then year 6, the effect diminishes to reach at maximum $5\%$ at the BAO scale and $10\%$ at $r=200~h^{-1}~\mathrm{Mpc}$. The predictions obtained from using the analytical $\xi_\ell$ in the integral constraint calculation for year 6 are shown in Fig. \ref{fig:ic} with the solid curves. Those predictions are perfectly in line with those obtained from taking the measured $\xi_\ell$, although on the largest scales, they provide more robust predictions as they are not affected by sample variance. In all cases, the effect remains well within the 1$\sigma$ statistical uncertainty expected at year 6, as shown with the green shaded area in Fig. \ref{fig:ic}. In future \Euclid cosmological analyses, this effect can be straightforwardly accounted for at likelihood level, when confronting measurements to model predictions, or simply neglected if the analysis is limited to scales below about $r=100~h^{-1}~\mathrm{Mpc}$.  

As a final remark, we note that we did not consider here the impact of the sample purity and completeness, as well as possible radial integral constraint effects \citep{demattia19}, which are beyond the scope of this paper and are the subject of forthcoming Euclid Collaboration papers.

\section{Conclusions}

This paper has described the development and validation of the processing function of the Science Ground Segment for estimating the 3-dimensional 2PCF, as part of the \Euclid mission. The methodology employed in the software is both robust and effective. The data handling process has been optimised through innovative spatial partitioning strategies, which drastically reduce computation time while maintaining high accuracy. These optimisations are crucial for meeting the \Euclid mission's stringent scientific requirement, which demand both speed and precision. This involves advanced algorithms such as kd-tree, octree, and linked-list, employed for pair counting and that ensure an efficient, yet exact, estimation of the 2PCF. The other key feature of the software is its optimisation and parallelisation capabilities. By leveraging the \texttt{OpenMP} application programming interface, the software achieves significant improvements in computational efficiency. These methods are indispensable for managing the extensive and complex data set anticipated from the \Euclid survey, which involves billions of galaxy pairs.

The software has undergone extensive validation using a variety of mock galaxy catalogues, which were designed to simulate the conditions and challenges of \Euclid survey data. Rigorous tests have demonstrated that the software can handle the anticipated data volume and complexity, ensuring that the \Euclid mission's scientific goals can be met. The validation process has confirmed the software's high accuracy and reliability in estimating the 2PCF, leading to an accuracy below that required by the \Euclid mission. Moreover, the software's integration into the \Euclid SGS pipeline exemplifies its readiness for large-scale deployment and its pivotal role in the mission's data processing framework for the production of highest-level scientific products for cosmology. This integration ensures that the vast amount of data collected by the \Euclid mission can be processed efficiently.

We have presented forecasts of 2PCF-GC performance in measuring the 2PCF at different stages of the Euclid Wide Survey. These forecasts are based on a realistic mock data set and illustrated the robustness and efficiency in estimating the 2PCF. This mock data set has also allowed us to investigate the impact of different pair line-of-sight definitions on the estimation of the multipole correlation function, highlighting differences in sensitivity to wide-angle effects over the scales that will be reached by \Euclid after 6 years of observations. We found that deviations in the monopole and quadrupole correlation functions are less than $5\%$ below $100\, h^{-1} \, {\rm Mpc}$, but can exceed $25\%$ at larger scales. We have also estimated the integral constraint bias affecting Euclid Wide Survey measurements. While this is expected to be minimal in the Euclid Wide Survey, due to the unprecedented precision required in the \Euclid measurements, the assessment of the magnitude and extent of the integral constraint effect remains critical. Using both measured and analytically modelled multipole correlation functions, we estimated that the integral constraint effect is of the order of $5\times 10^{-5}$, depending on the redshift interval and observing epoch considered. Our results indicate that the effect is most pronounced in the first year of the survey, reaching $10\%$ at the BAO scale and up to $30\%$ at $200\, h^{-1} \, {\rm Mpc}$, but diminishes in subsequent years. Nonetheless, future work will need to consider additional factors in the estimation of the 2PCF, such as sample purity, completeness, and potential radial integral constraint effects, which are beyond the scope of this paper. In particular, a first assessment of the impact of expected redshift errors on the 2PCF indicate that the results shown in this paper should be considered as optimistic (Risso et al., in prep.).

\begin{acknowledgements}

The authors would like to thank Jean-Charles Lambert for his help in using and extracting reliable computational measures from the LAM computer cluster. 
This work was supported by the ASI/INAF agreement n. 2018-23-HH.0 “Scientific activity for Euclid mission, Phase D”, the MIUR, PRIN 2017 research grant ‘From Darklight to DM: understanding the galaxy/matter connection to measure the Universe’ and the the INFN project “InDark”. FM acknowledges the financial contribution from the grant PRIN-MUR 2022 20227RNLY3 ‘The concordance cosmological model: stress-tests with galaxy clusters’ supported by Next Generation EU and from the grant ASI n. 2024-10-HH.0 ‘Attività scientifiche per la missione Euclid – fase E’.
This work has made use of CosmoHub. CosmoHub is developed and maintained by PIC, IFAE, CIEMAT, in collaboration with ICE-CSIC. It is partially financed by the European Union NextGenerationEU(PRTR-C17.I1) and by Generalitat de Catalunya, as well as by the grant EQC2021-007479-P funded by MCIN/AEI/10.13039/501100011033 and by the "European Union NextGenerationEU/PRTR".

\AckEC

\end{acknowledgements}

\bibliography{euclid.bib}

\label{LastPage}

\end{document}